\documentclass[11pt]{elsarticle}
\usepackage{amsmath}
\usepackage{subcaption}
\usepackage{caption}
\usepackage{hyperref}
\usepackage{comment}

\newcommand{\ii}{\mathrm{i}}

\title{
\vskip-1.5cm{\baselineskip14pt\rm
\centerline{\normalsize DESY 15-144\hfill ISSN 0418-9833}
\centerline{\normalsize HU-EP-15/37\hfill}
\centerline{\normalsize August 2015\hfill}}
\vskip1.5cm
Master integrals for the four-loop Sudakov form factor}

\author[HH]{Rutger H. Boels}
\author[HH]{Bernd A. Kniehl}
\author[Berlin,Beijing]{Gang Yang}

\address[HH]{II. Institut f\"ur Theoretische Physik, Universit\"at Hamburg\\ Luruper Chaussee 149, 22761 Hamburg, Germany}
\address[Berlin]{Institut f\"ur Physik, Humboldt Universit\"at zu Berlin \\ IRIS-Geb\"aude, Zum Gro\ss en Windkanal 6, 12489 Berlin, Germany}
\address[Beijing]{State Key Laboratory of Theoretical Physics, Institute of Theoretical Physics, \\Chinese Academy of Sciences, Beijing 100190, China}

\begin{document}

\begin{abstract}
The light-like cusp anomalous dimension is a universal function in the analysis of infrared divergences. In maximally ($\mathcal{N}=4$) supersymmetric Yang--Mills theory (SYM) in the planar limit, it is known, in principle, to all loop orders. The non-planar corrections are not known in any theory, with the first appearing at the four-loop order. The simplest quantity which contains this correction is the four-loop two-point form factor of the stress tensor multiplet. This form factor was largely obtained in integrand form in a previous work for  $\mathcal{N}=4$ SYM, up to a free parameter. In this work, a reduction of the appearing integrals obtained by solving integration-by-parts (IBP) identities using a modified version of {\tt Reduze} is reported. The form factor is shown to be independent of the remaining parameter at integrand level due to an intricate pattern of cancellations after IBP reduction. Moreover, two of the integral topologies vanish after reduction. The appearing master integrals are cross-checked using independent algebraic-geometry techniques explored in the {\tt Mint} package. The latter results provide the basis of master integrals applicable to generic form factors, including those in Quantum Chromodynamics. Discrepancies between explicitly solving the IBP relations and the MINT approach are highlighted. Remaining bottlenecks to completing the computation of the four-loop non-planar cusp anomalous dimension in $\mathcal{N}=4$ SYM and beyond are identified. 
\end{abstract}


\maketitle

\section{Introduction}

Maximally ($\mathcal{N}=4$) supersymmetric Yang--Mills theory (SYM) offers perhaps the best chance of truly solving an interacting four-dimensional quantum field theory. In addition, it is a proven stepping stone to pioneer new computational technology, as the large amount of supersymmetry renders many computations much simpler than their non-supersymmetric counterparts. In many cases, the resulting technology has transformed computational power in generic quantum field theories. For instance, in QCD, recent years have seen a boom of next-to-leading-order (NLO) computations considered unfeasible only a few years before, inspired ultimately by Witten's twistor string \cite{Witten:2003nn}. 

The focus of this paper is the computation of the {\it Sudakov} form factor, which is an observable that involves one gauge-invariant operator from the stress tensor multiplet and two on-shell massless states,
\begin{equation}
{\cal F}_2 = \langle p_1, p_2 | \mathcal{O} |0  \rangle \, .
\end{equation}
In $\mathcal{N}=4$ SYM, this form factor was first discussed and computed at the two-loop order in \cite{vanNeerven:1985ja}. It is noteworthy that the three-loop form factor was first computed in QCD \cite{Baikov:2009bg}, almost three years after the master integrals were found in \cite{Gehrmann:2006wg}. This result was then adapted to provide the three-loop answer in $\mathcal{N}=4$ SYM in \cite{Gehrmann:2011xn}. The latter computation cleanly shows that the computation within $\mathcal{N}=4$ SYM is technically much more straightforward, especially when employing modern unitarity-based methods instead of Feynman graph techniques. 

Since there is only a single scale in the problem of computing the Sudakov form factor, the dependence on this scale is fixed by dimensional analysis. Hence, the form factor evaluates to a function which only depends on the dimensional regularization parameter $\epsilon$ ($D=4-2 \epsilon$), the coupling constant and the number of colors $N_c$. Moreover, the dependence on $\epsilon$ is governed to a large extent by the known structure of infrared (IR) divergences. The divergent structure through $\frac{1}{\epsilon^2}$, for instance, is determined by the so-called cusp (or soft) anomalous dimension, which gets its name from its appearance in the computation of the light-like cusped Wilson line \cite{Korchemsky:1985xj, Korchemsky:1987wg}. The cusp anomalous dimension is a function which is universal for a given quantum field theory. 

In $\mathcal{N} =4$ SYM, an integral equation which determines the leading planar part of the cusp anomalous dimension in principle exactly was derived from integrability \cite{Beisert:2006ez}. This is to date the most powerful and precise manifestation of the cluster of ideas known as the AdS/CFT correspondence \cite{Maldacena:1997re}. However, much less is known about the non-planar part of the cusp anomalous dimension. At weak coupling, an immediate problem is that the first non-planar correction appears at four loops and has never been calculated to date, in any theory. Further motivation comes from a conjecture \cite{Becher:2009qa} that this correction may in fact vanish. This is based on the assumed completeness of the dipole contributions, which is consistent with the analysis of IR divergences in Yang-Mills theories explored through three-loop order, see also \cite{Gardi:2009qi}.
However, there is evidence from the Regge limit analysis in \cite{Caron-Huot:2013fea} that this naive dipole summation is not sufficient at four loops.  It is important to settle this issue by explicit computation. 

%This paper is part of the endeavor to compute the non-planar four-loop cusp anomalous dimension. 
%In our setup, the Sudakov form factor is to be computed in dimensional regularization and expanded to order $\frac{1}{\epsilon^2}$, down from the leading term at $\frac{1}{\epsilon^8}$. 
The integrand of the form factor was  obtained in \cite{Boels:2012ew} using color-kinematic duality \cite{Bern:2008qj, Bern:2010ue, Bern:2012uf} as an ansatz generator. The coefficients in the ansatz were fixed by a choice of unitarity cuts, up to one remaining parameter. 
%It was expected that this parameter may be fixed by imposing a completed set of unitarity cuts. 
To evaluate the integral, a major next step is to perform the integral reduction, in particular using integration-by-parts (IBP) reduction \cite{Chetyrkin:1981qh, Tkachov:1981wb}. This turns out to be highly non-trivial due to the complexity of four-loop integrals. The success of this reduction is a major achievement of this paper. In particular, this reduces the challenging non-planar cusp anomalous dimension problem down to the computation of an explicit set of master integrals. 

Based on the IBP reduction result, surprisingly, the free parameter in the integrand turns out to be truly free: it drops out of the full form factor after the reduction. This completes the determination of the integrand of the $\mathcal{N}=4$ SYM form factor at four loops. In other words, we prove that there is a one-parameter family of four-loop form factor integrands which satisfy color-kinematic duality. 
%It would be interesting to understand its physical implication. 
%, leaving the computation of the master integrals as . 

The master integrals are also explored independently by algebraic techniques. While the results are consistent with the IBP reduction in majority, some rare examples of mismatch present interesting open questions to understand further. The results from algebraic methods, with some qualifications, provide the master integrals of four-loop form factors in generic quantum field theory, including QCD.

The remaining problem of computing the four-loop form factor in $\mathcal{N}=4$ SYM is the evaluation of the master integrals. Many techniques are available to compute Feynman integrals, see \cite{Smirnov:2012gma} and references cited therein for an overview. However, the basis integrals of the four-loop form factor are rather non-trivial. They include non-planar four-loop integrals with up to quadratic irreducible numerators, the majority of which, to our knowledge, have never been integrated. We present partial numerical results and briefly survey the challenges involved in completing the computation, while leaving the full evaluation to future work.
%We found that most publicly available tools do not work on this set, although most do work for the three-loop form factor. 
%There seems to be an order of magnitude or more jump in complexity from three to four loops. 
%The main achievement of this article is to obtain a full basis of master integrals for the four-loop form factor in any quantum field theory.
% 

This article is structured as follows. First, the connection between form factor and cusp anomalous dimension is reviewed in section \ref{sec:setup}. This section also introduces some of the language to study Feynman integrals and discusses the issue of graph symmetry. 
%important issue of graph symmetry. 
In section \ref{sec:IBP}, we report on the IBP reduction of the Feynman integrals in the four-loop form factor in $\mathcal{N}=4$ SYM based on the {\tt Reduze} package \cite{vonManteuffel:2012np}. The independence of the sum on the free parameter reported in \cite{Boels:2012ew} is elucidated. We then proceed, in section \ref{sec:MINTish}, to study the basis of master integrals using algebraic techniques explored in \cite{Lee:2013hzt}. A brief survey of the numerical evaluation is given in section \ref{sec:towardsnumerics}. Appendices contain information on a three-loop cross-check and an alternative choice of basis integrals. In the supplementary material section, there are several ancillary files which contain results for the basis of master integrals. 
%These files should be considered as an integral part of this article.

\section{Review and setup} %overview}
\label{sec:setup}

In the following, extensive use will be made of the results in \cite{Boels:2012ew} for the four-loop form factor in $\mathcal{N}=4$ SYM, in particular of the graphics and tables in section $5$ of that paper. The numbering of the integral topologies refers to that paper. 

In the course of writing this paper, a typo was discovered in Table 5 of  \cite{Boels:2012ew}: the entry for the color factor of integral topology $26$ should be $24 N_c^2 \delta_{a_1 a_2}$. In particular, this integral topology does not contribute to the planar cusp anomalous dimension.

\subsection{Relation between form factor and cusp anomalous dimension}

The Sudakov form factor is an observable which here involves one gauge-invariant operator from the stress tensor multiplet and two on-shell massless states,
\begin{equation}
{\cal F}_2 = \langle p_1, p_2 | \mathcal{O} |0   \rangle \, .
\end{equation}
In the present paper, we focus on the four-loop Sudakov form factor in ${\cal N}=4$ SYM, for which the integrand was obtained in \cite{Boels:2012ew}.\footnote{In ${\cal N}=4$ SYM, form factors were first studied thirty years ago in \cite{vanNeerven:1985ja} and revived in the past few years at weak coupling \cite{Gehrmann:2011xn, Brandhuber:2010ad, Bork:2010wf, Brandhuber:2011tv, Bork:2011cj, Henn:2011by, Brandhuber:2012vm, Bork:2012tt, Engelund:2012re, Penante:2014sza, Brandhuber:2014ica, Bork:2014eqa} and at strong coupling \cite{Alday:2007he, Maldacena:2010kp, Gao:2013dza}. See also recent developments concerning form factors of non-Bogomolnyi--Prasad--Sommerfield (BPS) operators \cite{Wilhelm:2014qua, Nandan:2014oga, Loebbert:2015ova, Frassek:2015rka}.}

The Sudakov form factor for the stress tensor multiplet can be written as the tree-level form factor times a scalar function,
\begin{equation}
{\cal F}_2 = {\cal F}^{(0)}_2 \, {F_2} (g, N_c) \, .
\end{equation}
It is this function ${F_2}$ which will be computed. In momentum space, ${F_2}$ is a function of the only scale in the problem, $q^2 = (p_1 + p_2)^2$, the coupling constants, group theory Casimirs and $\epsilon$. Here, $q$ is the momentum associated with the gauge-invariant operator. Note that it does not obey a mass-shell condition, in contrast to the momenta of the gluons, $p_1^2 = p_2^2 = 0$.

The cusp anomalous dimension $\gamma_{\textrm{cusp}}$ is related to the two-point form factor through the universal exponentiation of IR divergences \cite{Mueller:1979ih, Collins:1980ih, Sen:1981sd, Magnea:1990zb}, which, in dimensional regularization, takes the form
\begin{equation}\label{eq:infraredexpo}
{\rm Log}[{F}_2] =  {\sum_l (- q^2)^{- l \epsilon} \frac{-g^{2l}\gamma_{\textrm{cusp}}^{(l)} }{4(l \epsilon)^2}  } + \mathcal{O}(\epsilon^{-1}) \, .
\end{equation}
The cusp anomalous dimension can be calculated as a perturbative series in the coupling constant. Displaying the dependence on group theory quantities up to fourth order, it reads
\begin{equation}
\gamma_{\textrm{cusp}} = \sum_{l} g^{2l} \gamma^{(l)}_{\textrm{cusp}} = a_1 g^2 C_A + a_2  g^4 C_A^2 + a_3  g^6 C_A^3 + g^8 \left( a_{4}^\mathrm{P}  C_A^4 + a_{4}^\mathrm{NP} d_{44} \right) +  {\cal O}(g^{10}) \, ,
\end{equation}
where $C_A$ is the Casimir which, for the SU($N_c$) gauge group, simply has value $N_c$. The quantity $d_{44}$ is a particular group theory Casimir invariant which, for SU($N_c$), has the value $N_c^4 + 36 N_c^2$. In previous works, the constants $a_1$, $a_2$, $a_3$ and $a_4^P$ were calculated in perturbation theory \cite{Korchemsky:1987wg, Bern:2005iz, Bern:2006ew, Cachazo:2006az, Henn:2013wfa}\footnote{The three-loop cusp anomalous dimension in ${\cal N}=4$ SYM was first obtained (conjecturally) in \cite{Kotikov:2004er} by extracting the leading transcendentality part of the QCD result \cite{Moch:2004pa, Vogt:2004mw}. In this case, the cusp anomalous dimension is related to the anomalous dimension of the twist-two operator in the limit of spin $j\rightarrow \infty$.} and using integrability \cite{Beisert:2006ez}. In 't~Hooft's planar limit \cite{'tHooft:1973jz}, $N_c \rightarrow \infty$ with $g^2 N_c$ fixed, it is clear that the first subleading color correction to the cusp anomalous dimension occurs at four loops. 
%See the appendix of \cite{Boels:2012ew} for a group theory analysis to eight loops. 

%Compared to the leading corrections in the planar limit, virtually nothing is known about the non-planar corrections. 
%The first non-planar correction to the cusp anomalous dimension is the ultimate goal of the research presented here. 
From the general exponentiation formula in equation (\ref{eq:infraredexpo}), this non-planar correction can be isolated from the non-planar part of the form-factor at four-loop order at order $\frac{1}{\epsilon^2}$ in the $\epsilon$ expansion. Since the integrals present in the form factor naively diverge as $\frac{1}{\epsilon^8}$, this gives an important check on any computation of the non-planar form factor: the first six orders in the $\frac{1}{\epsilon}$ expansion must vanish. 

The planar correction to the cusp anomalous dimension at four loops can be expressed as a sum of rational coefficients times either $\zeta(3)^2$ or $\pi^6$ \cite{Bern:2006ew, Cachazo:2006az, Henn:2013wfa}, which are both constants of {\it transcendentality} six. Based on the maximal-transcendentality principle \cite{Kotikov:2004er}, it is natural to conjecture that the four-loop non-planar correction $a_4^\mathrm{NP}$ is given by a combination of the same transcendental numbers, if not vanishing.
%at least of universal transcendentally, most likely of the same order as the planar four-loop correction.

\subsection{Feynman integrals}

%Feynman integrals are central objects in the explicit computation of quantum effects in quantum field theory. 
An $L$-loop Feynman integral with $n$ so-called indices $a_1, \ldots, a_n$ is an integral which can be written as
\begin{equation}\label{eq:deffeynmanint}
I(a_1, \ldots, a_n) \equiv \int d^D l_1 \ldots d^Dl_L  \left(1/D_1\right)^{a_1} \ldots  \left(1/D_n\right)^{a_n} ,
\end{equation}
where $D_i$ are inverse propagators. 
%, that is, expressions which are the invariant length squared of a linear sum of loop and external momenta. 
For future reference, we define the absolute value of the sum over positive and negative indices to be $t$ and $s$ respectively. The parameter $s$ is the numerator power and $t$ is the propagator power. 
%Note that the number of different propagators is smaller than or equal to $t$. 
In this article, the indices are strictly integer valued. Integrals which have the same non-zero indices for propagators are said to belong to the same {\it sector}. In the special case where all propagator powers are one, the integral is referred to as the {\it corner integral} of the sector. 

%The form of the propagators depends on the kinematics of the problem under study: important is the number of possible independent quadratic expressions. For four-loop form factors with three external momenta (two after momentum conservation) there are eighteen independent inner products between loop and external momenta which involve at least one loop momentum. Hence, one can pick a basis of eighteen propagator-type factors. Once a basis has been chosen, any polynomial of the momentum invariants can uniquely be expressed in terms of this basis, up to terms which are proportional to $p_1^2$, $p_2^2$ and $(p_1+p_2)^2$,\footnote{In the problem under study, $p_1^2 = p_2^2 = 0$. As a trick to compute the polynomial decomposition without introducing Lorentz vectors, first add $p_1^2$ and $p_2^2$ to the basis and consider equation \ref{eq:decomp} for each momentum invariant in the problem separately. Since the coefficients $\alpha_i$ are, for the application of interest here, at most rational numbers, this can be solved by treating the Lorentz vectors as integers. Then, one evaluates this equation $21$ times for large random integer values and inverts the resulting matrix system exactly by computer algebra.}
%\begin{equation}\label{eq:decomp}
%(\textrm{polynomial}) = f\left(\{D_i\}, p_1^2 , p_2^2,  (p_1+p_2)^2\right) \,.
%\end{equation}

The problem under study in this article consists of a series of four-loop twelve-propagator integrals. For each integral topology, one can construct a basis by starting with a parametrization of the loop momenta of the graphs. Then, one needs to pick six additional propagators to get a complete set. These additional propagators are important for expressing irreducible numerators. Given a numerator, one can express this uniquely into the basis. The choice of numerators is by nature somewhat arbitrary. If one focuses solely on additional propagators which are products of differences of two momenta, then this can be done by trial and error. A more systematic approach is needed, if the aim is to include graph symmetry. 

As an example, let us consider integral topology $26$. Its propagators can be parametrized as:
\begin{multline}\label{eq:propstop26}
l_6^2, \, l_5^2,\, l_4^2\,, l_3^2\,,(-l_4 + l_5)^2\,,(l_3 - l_4)^2,\,(-l_5 + p_1)^2,\,(-l_6 + p_2)^2, \\
(-l_4 + l_5 + l_6)^2,\,(-l_3 + p_1 + p_2)^2,\,(l_3 - l_4 + l_6 - p_2)^2,\, (-l_3 + l_4 - l_5 - l_6 + p_1 + p_2)^2 \,.
\end{multline}
This topology has two independent permutation symmetries, as will be discussed later. It can be checked that the following six additional propagators parametrize all irreducible numerators:
\begin{equation}\label{eq:top26addpropswosym}
(l_3-l_5)^2 \, , \ \ 
(l_3-l_6)^2 \, , \ \ 
(l_5-l_6)^2 \, , \ \ 
(l_4-p_1)^2 \, , \ \ 
(l_4-p_2)^2 \, , \ \ 
(l_5-p_2)^2 \, .
\end{equation} 
In section \ref{app:basis}, choices of propagators are explicitly given for all $34$ topologies.

A useful step toward computing the Feynman integrals is the so-called $\alpha$ representation, see e.g.\ \cite{Smirnov:2012gma}. 
%Basically, one uses a more elaborate version of the usual Feynman trick of combining numerators in a loop integral. 
The result is an integral over $\alpha$ parameters, one for each propagator,
\begin{equation}\label{eq:defalphaparam}
I(a_1, \ldots, a_n) \propto  \frac{1}{\prod_i \Gamma[a_i]} \int_0^{\infty} d\alpha_1 \ldots d\alpha_n \prod_i \alpha_i^{a_i -1} U^{-D/2} e^{- \ii F / U} \, .
\end{equation}
where the normalization constant, which is unimportant for our purposes, has been suppressed. $U$ and $F$ are certain polynomials of the $\alpha$ parameters of homogeneity $L$ and $L+1$, respectively. See \cite{Smirnov:2012gma} for further discussion and \cite{asmirnovwebsite} for a {\tt Mathematica} code to compute them from given propagators. As is customary, the labels on the $\alpha_i$ parameters correspond to the position in a given ordered list of propagators, such as the one in equation (\ref{eq:propstop26}) for integral topology $26$. 
Equation (\ref{eq:defalphaparam}) can cover negative indices, that is, numerators as well. %For this, one observes that the Gamma function prefactors tend to push the total result to zero in the limit in which the indices become negative integers. Hence, a non-zero result is only obtained if the $\alpha$ parameter integral diverges as $\propto\frac{1}{\alpha_i}$ if the index to be taken negative is $a_i$. Inspection of the integrand gives as an obvious source of divergences the integration region $\alpha_i \approx 0$. Laurent expanding the integrand about this point and integrating gives a definite formula for Feynman integrals with numerators, under the assumption that there are no other sources of divergences. This assumption is made in the computer code {\tt FIESTA} \cite{Smirnov:2013eza}, for instance. 

\subsection{Graph symmetry}

Several of the graphs in the set under study have a graph symmetry: the graphs are mapped to themselves under a permutation of some of the edges and external legs. Exploiting graph symmetry can be of great help in simplifying and cross-checking computations if it is present. 

In \cite{Boels:2012ew}, graph symmetry was indicated briefly by the symmetry factor. These symmetries can be checked or found by using the built-in capabilities of {\tt Mathematica}. %\footnote{{\tt Mathematica~10} can handle multi-graphs. When importing .dot files, though, both propagators of a bubble subgraph will end up with the same label, and this needs to be corrected by hand.} 
These permutations can include permutations of the external legs. An alternative strategy is to compute the $U$ and $F$ polynomials of the $\alpha$ parameter integral, equation (\ref{eq:deffeynmanint}). Then, one can simply check all possible permutations and see if they leave these polynomials invariant. Due to the factorial growth of this problem, the feasibility of this analysis strongly depends on the algorithm: a naive implementation ran out of steam for ten propagators. A faster algorithm is presented in \cite{Pak:2011xt}. 

A symmetry of the graph of an integral can correspond to a symmetry of a corresponding Feynman integral. For a given parametrization of the loop momenta of the graph, a permutation symmetry of the graph may be undone by a shift of the loop momenta, combined with a permutation of the external legs. If the latter leaves the integral invariant, the result is a symmetry of the Feynman integral.  

For the example of topology $26$, the two independent permutations are generated from
\begin{eqnarray}\label{eq:premium}
\begin{array}{cl} 
l_6 & \leftrightarrow l_5 \,, \\  
l_4 & \rightarrow -l_4 + l_5 + l_6 \,, \\ 
l_3 & \rightarrow  -l_3 + p_1 + p_2 \,, \\ 
p_1 & \leftrightarrow p_2 \,,
\end{array} 
& \qquad {\rm and} \qquad \ \  & 
\begin{array}{cl} 
l_6 & \rightarrow -l_5 + p_1 \,, \\   
l_5 & \rightarrow -l_6 + p_2 \,, \\ 
l_4 & \rightarrow l_3 - l_4 \,, \\ 
p_1 & \leftrightarrow p_2 \,,
\end{array}
\end{eqnarray}
as can be checked explicitly in equation (\ref{eq:propstop26}). Since the integrals, by dimensional analysis, only depend on $(p_1+p_2)^2$, they are left invariant by the permutation of $p_1$ and $p_2$. 
%The latter check is important, as, for instance, the following map also corresponds to a symmetry of the edges:
%\begin{equation}
%\begin{array}{cl} 
%p_2 & \rightarrow -p_1 - p_2 \,, \\ 
%l_6  & \leftrightarrow -l_3 \,, \\  
%l_4  & \rightarrow -l_4 + l_5 \,,
%\end{array}
%\end{equation}
%while this is {\it not} a symmetry of the Feynman integral. In general, the set of integral topologies contains surprisingly many graph symmetries which are not integral symmetries. These all involve a permutation of the operator leg with an on-shell-state leg which sends $q^2 \rightarrow - q^2$. Hence, these would-be {\it graph symmetries} have to be excluded. With this exclusion, all graph permutation symmetries correspond to symmetries of the Feynman integrals. In the following, those graph symmetries which are also symmetries of the Feynman integrals will be referred to as {\it permutation symmetries}.

The momentum map given above corresponds to a simple permutation pattern for a given list of propagators. Let us consider, for instance, the scalar integral in topology $26$, with the propagators as listed in equation (\ref{eq:propstop26}). The two independent permutation symmetries of the graph correspond to the following cycles:
\begin{equation}
\label{eq:symmetry-top26}
\begin{array}{c} 
\{ \{ 1, 7 \}, \{ 2, 8 \}, \{ 3, 6 \}, \{ 5, 11 \}, \{ 9, 12 \} \} \,, \\
 \{ \{ 1, 2 \}, \{ 3, 9 \}, \{ 4, 10 \}, \{ 6, 12 \}, \{ 7, 8 \} \} \,.
\end{array}
\end{equation}
Together with the trivial permutation and their product, these form a four-element representation of the permutation group. A given propagator can be mapped to other propagators in the set, but only if they are in the orbit of the given permutation. The orbits of the above permutation group are
\begin{equation}
\{\{1,2,7,8\},\{3,6,9,12\},\{4,10\},\{5,11\}\} \, .
\label{eq:orbits-group-26}
 \end{equation}
This can be used to simplify the sector decomposition method employed in {\tt FIESTA}, for instance, as will be shown in section \ref{subsec:numerical-details}.

Given a permutation symmetry of a particular topology, ideally, additional propagators would be found which yield a complete basis of propagators and also incorporate the permutation symmetry. One can search for symmetric numerators systematically. Permutation symmetries either leave propagators invariant or interchange two propagators. Hence, the six propagators to be added to the set contain either zero, one, two or three pairs of interchanges and, consequently, six, four, two or zero invariant propagators. Starting from a generic linear combination of all four loop momenta and two external momenta, one can construct the most general pair combination as well as the most general invariant propagator in terms of a number of free variables. Since propagators are only determined up to a sign by the appearance of the square, there are two times two possible (pairs of) polynomials which come out of this. Hence, all possible appearances of the permutation symmetry can be parametrized. 

Then, one checks if the resulting set of eighteen propagators is linearly independent. In several cases (e.g.\ topology $20$), no set exists at all. In addition, there are integrals with multiple permutation symmetries which do not permit a choice of six additional propagators with all permutation symmetries manifest. In cases in which there are sets of numerators with explicit permutation symmetries, typically many free variables exist. These may then be fixed by hand aiming for propagators as simple as possible. Care should be exercised not to choose parameters such that accidental linear dependencies in the total set of propagators are introduced. 

%This property is needed if one wants to use symmetry as an input in the {\tt Reduze} code or have explicit symmetry properties.
For the example of topology $26$, the momentum maps in equation (\ref{eq:premium}) do not map the additional numerators in equation (\ref{eq:top26addpropswosym}) into a permutation of themselves. The following set of additional numerators has explicit permutation symmetry: 
\begin{equation}\label{eq:top26addpropssym}
\begin{array}{l}
(l_3 + 2 l_5 - 3 p_1)^2 \,, \\
(-l_3 + 2 l_6 + p_2)^2 \,, \\ 
(l_3 - 2 l_6 - p_1 + 2 p_2)^2 \,,
\end{array} 
\qquad
\begin{array}{l}
(l_3 + 2 l_5 - p_2)^2 \,, \\
(l_3 - l_4 + l_5 - p_2)^2 \,, \\
(l_4 - l_6 - p_1 + p_2)^2 \,.
\end{array}
\end{equation} 
This example shows a general principle: the price of manifest permutation symmetry can be a much more complicated numerator structure. It would be very interesting to fix guiding principles for choosing sets of numerators. 

The considerations above result in two sets of propagators. For each integral topology, there is a choice of eighteen propagators which can be used to express any integral in this particular topology. In particular, any numerator can be expressed in terms of the basis. In the first set, the aim was to include numerators as simple as possible, while, in the second, permutation symmetries were taken into account as much as possible. Obtaining the symmetric set is the result of a fairly lengthy computation. The simplest choice in the first set is listed in section (\ref{app:basis}). All results reported here were obtained with this set, unless explicitly mentioned otherwise. 

%Graph symmetry and the symmetry of the numerators are an example of an issue which is relatively trivial at three loops, but becomes involved at four loops. 

\section{IBP reduction and its output}\label{sec:IBP}

The set of Feynman integrals in equation (\ref{eq:deffeynmanint}) is over-complete, as there are relations between different Feynman integrals. A particular example are the IBP relations \cite{Chetyrkin:1981qh, Tkachov:1981wb}, which follow from 
\begin{equation}
0 = \int d^D l_1 \ldots d^Dl_L  \,\, \frac{\partial}{\partial l_i^{\mu}} \, X \,.
\end{equation}
Working out the right-hand side leads to a linear relation of different Feynman integrals. By solving a system of such equations, one may express a general Feynman integral in terms of some basic integrals. 
%
%Working out the right-hand side leads to a sum over integrals which can all be expressed in terms of the basis of the diagram topology, if the free index is contracted with either one of the loop momenta or one of the external momenta. 
%In our four-loop problem, there are six relations for each of the four loop momenta for each choice of $X$, for each topology. 
%Hence, there are very many relations, many of which are furthermore dependent.

A different way of phrasing the problem is to envision the system of IBP relations as a large matrix equation, with the integrals combined into a vector. The standard way of solving a problem of this type is Gaussian elimination. However, the output of this algorithm for a non-invertible matrix problem, such as the one under study here, depends on the ordering of the integrals in the vector. This is the essence of Laporta's algorithm for IBP reduction \cite{Laporta:2001dd}: one picks an ordering of the integrals. This ordering should be such that complicated integrals are expressed in terms of simpler ones in general. For instance, integrals with smaller values of the parameter $s$ (measuring numerator powers) are preferred, as are integrals with less numbers of different propagators. In addition, smaller values of the parameter $t$ (measuring propagator powers) are to be preferred. Within these general choices, in practice, the exact criteria can differ between two different implementations of Laporta's algorithm. The results of these two implementations are, therefore, in general not the same, but related by a change of basis. 

Given a system of equations, the output of Laporta's algorithm is a reduction of all integrals in a given set down to integrals which cannot be further reduced from the given equations using the given ordering. The left-over integrals are known as \emph{master integrals}. These integrals depend on the given set of equations, although typically, if the given set of equations is {\it large} enough, the set of master integrals tends to converge. 
%Recently, a method to obtain the master integrals alternative to solving the IBP relations was proposed in \cite{Lee:2013hzt} and implemented in the {\tt Mint} package.  

Various private and public implementations of Laporta's algorithm exist, such as {\tt AIR} \cite{Anastasiou:2004vj}, {\tt FIRE} \cite{Smirnov:2008iw, Smirnov:2013dia, Smirnov:2014hma} and {\tt Reduze} \cite{vonManteuffel:2012np,2010CoPhC.181.1293S}.  See {\tt LiteRed} \cite{Lee:2012cn, Lee:2013mka} for an alternative approach to IBP reduction. We explored  {\tt FIRE}, {\tt Reduze} and {\tt LiteRed} in some detail for the four-loop form factor problem. 
%to evaluate which one could solve the IBP reduction for the form factor in $\mathcal{N}=4$ SYM. 
Only {\tt Reduze} succeeded in solving the problem, probably due to its better parallelisation implementation. %The problem with the four-loop form factor is that it is at the edge of current technology. 

\subsection{Relevant implementation details}

%Various private and public implementations of Laporta's algorithm exist, such as {\tt AIR} \cite{Anastasiou:2004vj}, {\tt FIRE} \cite{Smirnov:2008iw, Smirnov:2013dia, Smirnov:2014hma} and {\tt Reduze} \cite{2010CoPhC.181.1293S, vonManteuffel:2012np}.  See {\tt LiteRed} \cite{Lee:2012cn, Lee:2013mka} for an alternative approach to IBP reduction. The problem with the four-loop form factor is that it is at the edge of current technology. 
%We explored both {\tt FIRE} and {\tt Reduze} in some detail to evaluate which one could solve the IBP reduction for the form factor in $\mathcal{N}=4$ SYM. {\tt Reduze} was able to solve this.

{\tt Reduze} is designed to run in massive parallel mode on a cluster. As an input, it takes a family of integrals. This family is treated as an ordered set: lower integrals in the set are considered simpler. Integrals are reduced as far as possible to lower integrals. This cuts down on the computational workload considerably, especially in a large-scale problem such as the one under study here. As parameters, {\tt Reduze} takes a range of propagator powers (the $t$ variable) as well as a range of numerator powers (the $s$ variable) and constructs out of these all possible IBP relations for a given sector. These are then solved sector by sector, starting from the simplest sector, which is the bottom sector of the first member of the integral family. The number of relations tends to grow very quickly along either the $t$ or the $s$ axis.

A problem we encountered in the public version of {\tt Reduze} in parallel mode was that it tended to crash when handling many large files on the cluster file system during the identity generation stage. Due to this problem it is unfeasible to obtain an IBP reduction with the public version of {\tt Reduze}, and a fix is needed. The problem was that a file was reported as not present in the file system, while a manual check thereafter did uncover this particular file. This is due to the internal file handling structure of {\tt Reduze},\footnote{We thank A. von Manteuffel for explaining this.} where all processes are allowed to read from and write to disc. As a work-around, the program can be forced\footnote{This can be done in the function \texttt{set\_job\_status} in the file \texttt{jobqueue.cpp}.} to sleep for a time-out of fifty seconds, followed by a $500$-second time-out if the file is still not found. This two-stage timeout resolved the crashing problem due to disk space usage. The loss of productivity for a few minutes is a small price to pay to prevent the total collapse of the computation.

During the running of the four-loop reduction, {\tt Reduze} still crashed occasionally when handling integrals in sectors with many, typically twelve or eleven propagators. These integrals require a large memory usage each, which can easily overwhelm one of the nodes of the cluster. Sometimes this can be due to bad scheduling of processes, with all {\it root} processes which consume most memory running on a single node; this situation can somewhat be avoided at start-up by passing scheduling instructions to the message-passing-interface (MPI) program.\footnote{One uses {\tt -bynode} in {\tt openmpi}.} By contrast, for lower propagator count, the limiting factor for speed is mostly CPU time. These two situations require opposite numbers of allowed parallel processes, which in turn requires occasional input by the user during running. It should be noted that, although {\tt Reduze} works in parallel, there is a limit to the number of cores which are assigned to a single process while still increasing computational speed. This saturation is sector dependent, but certainly under a hundred cores. 

The chosen IBP relations are in the range $t \leq 12$ and $0\leq s \leq 2$, with some extensions to $t=13$ for specific integral topologies. Since {\tt Reduze} features a choice of used relations, there can be problems with the reduction at the outer {\it edges} of these relations: there can be unresolved integrals. A simple check of this is to observe the file size of the reductions in a sector and compare it to the average for other integral topologies in a sector with the same numbers of propagators. A sudden increase in file size tends to indicate unreduced integrals: the unreduced integrals appear with much more complicated prefactors than the others, typically involving polynomial ratios with polynomial orders an order of magnitude above the norm. This is more than just a nuisance. The abnormal prefactors tend to be highly divergent in the limit $\epsilon \rightarrow 0$, typically $\propto\frac{1}{\epsilon^{15}}$ or worse. Since the form factor itself is expected to diverge at worst as $\frac{1}{\epsilon^{8}}$, this implies massive cancellations among the integrals. This particular problem occurred in the basis of section \ref{app:basis} in integral topologies $20$ and $26$, for integrals with at most ten propagators. This problem was resolved using consistency conditions, see two paragraphs down. 

We reduced using the set of numerators in section \ref{app:basis}. A secondary computation using numerators with explicit permutation symmetries failed; at a certain eleven-propagator integral, the MPI protocol ran into an issue with maximal size of a single message, which is set by the maximum size of an integer on a $32$-bit machine, see \cite{Hammond:2014:IBE:2690883.2690884} for an explanation of the problem and a possible resolution. This is a result of a choice of protocol and not related to the physical architecture used. In principle, this problem could be circumvented by rewriting the {\tt Reduze} code to handle messages of this size. 

Since the set of numerators with the simplest structures was used, integral relations from permutation symmetries are not explicit. This allowed a simple check on the results: take a graph symmetry relation and use the IBP reduction. If {\tt Reduze} found the relation, the result must be zero. We checked permutation symmetry on maximal-propagator integrals with up-to-quadratic numerators. Surprisingly, one single additional relation was found which involved exactly all master integrals with abnormal prefactors. Plugging this relation into the reduction eliminated the abnormal prefactors, reducing them to much more normal-looking ones.

\subsection{Results}

The computation led to a result after several months of computation on a fairly large cluster.\footnote{Memory usage on single sectors can exceed 200 GB. More detailed information is available on request.} The complete list of the master integrals obtained with this reduction is attached, both in the planar and non-planar sectors, as well as a sample result for topology $25$. Overall, the integrals which appear after reduction are simpler. There are two integrals with two numerator powers. One of these contains only eight propagators and is easy to integrate numerically. The other, however, has the full twelve propagators and occurs in integral topology $26$. It turns out that this particular integral can be integrated numerically (see section \ref{subsec:numerical-details}). 

Any basis of master integrals listed is arbitrary to some extent. For a given sector, an obvious first choice for a master integral is its corner integral. Beyond this, there is a choice of a (typically single) numerator or doubled-up propagator. The statistics in Table \ref{table:counting_Reduze} relates to the direct output of {\tt Reduze}, taking into account the one extra relation found from permutation symmetry. The statistics is split into those integrals which contribute to the planar and non-planar parts of the form factor. 

\begin{table}[ht]

\caption{Master integral statistics of obtained IBP reduction.}
\begin{subtable}{.5 \linewidth}
\caption{planar form factor} 
\centering
\begin{tabular}{c | c c c} % centered columns (4 columns)
\hline %inserts double horizontal lines
\# props & $s=0$ & s=1 &$ s=2$ \\ [0.5ex] % inserts table
%heading
\hline % inserts single horizontal line
12 & 8 & 6 & 0 \\ % inserting body of the table
11 & 18 & 2 & 0  \\
10 & 43 & 9 & 0 \\
9  & 49 & 1 & 0 \\
8  & 51 & 4  & 1  \\
7 & 25 & 0 & 0 \\
6 & 8 & 0 & 0 \\
5 & 0 & 0 & 0  \\ 
\hline %inserts single line
sum & 203 & 22 & 1  \\ [1ex] % [1ex] adds vertical space
\end{tabular}
\end{subtable}%
\begin{subtable}{.5\linewidth}
\caption{non-planar form factor} 
\centering
\begin{tabular}{c | c c c} % centered columns (4 columns)
\hline %inserts double horizontal lines
\# props & $s=0$ & s=1 &$ s=2$ \\ [0.5ex] % inserts table
%heading
\hline % inserts single horizontal line
12 & 10 & 10 & 1 \\ % inserting body of the table
11 & 13 & 3 & 0  \\
10 & 34 & 10 & 0 \\
9  & 29 & 1 & 0 \\
8  & 32 & 3  & 1  \\
7 & 13 & 0 & 0\\
6 & 7 & 0 & 0 \\
5 & 1 & 0 & 0  \\ 
\hline %inserts single line
sum & 139 & 27 & 2  \\ [1ex] % [1ex] adds vertical space
\end{tabular}

\end{subtable}
\label{table:counting_Reduze}
\end{table} 

In these tables, there is quite some double counting of integrals which appear both in the planar and non-planar sectors. Using the known result for the sum of the integrals in the planar sector, one can eliminate several of the integrals in the non-planar sector. This is more efficient than might be expected, as there are four integral topologies ($21$, $25$, $30$ and $31$) which each appear with the {\it same} overall color factor, $ 2 N_c^4 + 24 N_c^2$. This can be used to eliminate these integral topologies from the non-planar sector in favor of the integrals in the purely planar sector and the known planar-form-factor result. The resulting sum contains more integrals ($260$ in total) than the non-planar sector by itself. 

The choice of master integrals in the provided results is taken from the output of {\tt Reduze}. It is possible to choose other sets of master integrals, in particular the ones with numerators, to aim at simplifying, for instance, the prefactors of the integrals in the form factor or to make use of a permutation symmetry of the scalar skeleton of the integral. In particular, one aim could be to make these as weakly divergent as possible in the limit $\epsilon \rightarrow 0$. The power of divergence in this limit determines the order to which the integral is to be expanded to get the cusp anomalous dimension.

\subsubsection*{Cross-check: Multiple reductions}

The IBP reduction was run twice, with a second reduction involving a reshuffling of the integral topologies to put planar topologies first. It was checked that the two reductions commute: reducing with the second reduction followed by the first gives the same result as reducing with the first, after taking into account the extra relation found from permutation symmetries. Before taking into account this relation, there are seeming mismatches between several of the integrals which appear in the full four-loop form factor: these mismatches are all proportional to the extra relation. This is a consistency check for the IBP reduction. 

%{\tt FIRE}5 was tried for integral topologies $17$, $25$ and $26$, using the integrals appearing in the form factor result as input. However, for all of these, the program crashed after several days to weeks of computation. An autopsy is under way. 

\subsection{Cancellations in the form factor after IBP reduction}

The IBP reduction displays two separate patterns of interesting cancellations for the form factor result: one pattern takes place within the separate integral topologies, while the second is between different integral topologies. 

The first pattern is that, for integral topologies $18$ and $20$, the contributions to the form factor vanish after IBP reduction. Generically, the Feynman integrals in these topologies do not vanish; there are twelve-propagator master integrals. However, the specific combination of the integrals in the $\mathcal{N}=4$ SYM form factor reduce to zero. Considering the complicated numerators for topologies $18$ and $20$, this is rather nontrivial and not obvious at all at the integrand level. It should be noted that these integrals already do not contribute to the form factor: their color factors are also zero. Such cancellations, however, do not happen for the other four zero-color-factor topologies, $8$, $11$, $15$ and $16$, whose integrals reduce to nontrivial results. It would be very interesting to find an explanation for these vanishing integrals, as, in physics, a zero almost always has a physical and important symmetry explanation. 

Another intriguing pattern emerges when tracking the dependence on the single free parameter in the form factor integrand which was left in \cite{Boels:2012ew}. There, the free parameter in the form factor multiplies a combination of $24$ out of $34$ different integral topologies which obeys all the physical constraints imposed in \cite{Boels:2012ew} separately. In particular, this combination is color-kinematic dual by itself.  One may expect that, by applying other, more complicated cuts which are not exploited in \cite{Boels:2012ew}, this parameter may be fixed uniquely. The surprise is that, when summed with the correct color and symmetry factors (and only then), the dependence on the free parameter drops out of the form factor after IBP reduction. This means that the color-kinematic-dual representation of the four-loop form factor contains a truly free parameter!

This detailed cancellation 
%is highly nontrivial. It 
involves an order of fifty different master integrals with up to ten propagators and occurs in both the planar and non-planar sectors. Most of the cancellations of the coefficients of the master integrals involve integrals out of all of the eighteen integral topologies which involve the free parameter and have nontrivial color factors. 
%As an example of a simple cancellation, let us consider the master integral
%\begin{equation}
%\textrm{INT}[1, \{0, 1, 1, 0, 1, 1, 0, 1, 1, 0, 1, 0, 0, 1, 0, 1, 1, 0\}]\, ,
%\end{equation} 
%which occurs in the reduction of topologies $13$, $16$ and $17$. Topology $16$ has vanishing color factor, and the symmetry and color factors of the other two are the same. The coefficient of this particular master integral after IBP reduction is $-\frac{3 \epsilon^2}{7 (1 + 4 \epsilon)^2}$ in sector $13$ and one of the opposite sign in sector $17$. A more intricate 
As an intricate example, consider the coefficient of the master integral 
\begin{equation}
\textrm{INT}[2, \{-1, 0, 1, 1, 1, 1, 1, 0, 0, 0, 1, 1, 0, 0, 1, 0, 0, 0\}] \,,
\end{equation}
which occurs in topologies $12$, $14$, $15$, $17$ and $19$. The coefficient in topology $12$ reads:
\begin{equation}
 \frac{-242 + 2861 \epsilon - 4906 \epsilon^2 - 37519 \epsilon^3 + 133706 \epsilon^4 - 
 18520 \epsilon^5 - 378200 \epsilon^6 + 
 366000 \epsilon^7}{21 \epsilon^4 (1 + 4 \epsilon) (-1 + 5 \epsilon) (1 + 5 \epsilon)}\, ,
\end{equation}
which is exactly $-2$ times the coefficient of this master integral in topology $19$. Taking into account the symmetry factors of topologies $12$ and $19$, which are $2$ and $1$, respectively, this leads to a cancellation. Similarly, the coefficients of this master integral in topology $14$ cancels against that in topology $17$, while the color factor of topology $15$ is zero. The other master integrals typically involve much more intricate cancellations. 

It would be very interesting to reach a better understanding of the uncovered cancellations. On a practical level, this pattern completes the determination of the form factor integrand \cite{Boels:2012ew}. To our knowledge, this is the first example of a color-kinematic-dual integrand representation that contains a free parameter. It would be fascinating to see if there is any deeper physical meaning of the free parameter, for example, as due to some gauge symmetries. This free parameter can be set to any value at the start of the computation. A natural value is $-1$, which simplifies the naive appearance of the integrals. This cancellation provides strong evidence for consistency of both the integrand result in \cite{Boels:2012ew} and the IBP reduction performed in the present paper.

\section{Counting master integrals using algebraic techniques}\label{sec:MINTish}

%As studied in the previous section, IBP reduction allows the reduction of a given integrand to a linear combination of a set of master integrals.
%According to Laporta's algorithm \cite{Laporta:2001dd}, one generates IBP identities for integrals up to the maximum values of $t$ and $s$, corresponding to the powers of propagators and numerators, respectively. The master integrals obtained in such a way are thus limited to the given input values of $t$ and $s$, and, therefore, only apply to a particular problem. 

In \cite{Lee:2013hzt}, Lee and Pomeransky developed a method to count the number of master integrals using a technique adapted from algebraic geometry. This is partially based on previous work by Baikov \cite{Baikov:2005nv}. This method only relies on the topology of the integral and, therefore, applies for arbitrary values of $t$ and $s$. In particular, it is independent of obtaining an explicit IBP reduction and, hence, of the results of the previous section.  This idea was also implemented in a {\tt Mathematica} package called {\tt Mint} \cite{Lee:2013hzt}. In this section, we develop and apply this method to count the number of master integrals of four-loop form factors. 
%First, the main idea will be reviewed. Then, we focus on its application to the set of four-loop integrals under study here. Some limitations of the used technology as well as ways how to deal with them are discussed.

Similarly to the start of IBP reduction, one first needs to determine the unique physical sectors for a given integral topology. As described in \cite{Lee:2013hzt}, it is convenient to perform this step using {\tt LiteRed} \cite{Lee:2012cn, Lee:2013mka}. In particular, {\tt LiteRed} determines the sector symmetries, which can be taken into account automatically in {\tt Mint}. In the next step, the counting of master integrals is done independently for each physical sector.

Let us consider a physical sector of $L$-loop topology and $m$ propagators. From the associated polynomials $U$ and $F$ in equation (\ref{eq:defalphaparam}), one can define a new polynomial\footnote{Equivalently, one can consider a different polynomial using Baikov's representation as discussed in \cite{Lee:2013hzt,Baikov:2005nv}. For all cases being checked, the two polynomials always give the same result, while the use of the polynomial $G$ is usually much more efficient in the computation.}
\begin{equation}
G(\vec\alpha) = U(\vec\alpha) + F(\vec\alpha) \,,
\end{equation} 
which has polynomial degree $L+1$. Using the duality of homology and cohomology (see \cite{Lee:2013hzt} for more details), the number of master integrals is related to the number of independent integral contours. The latter depends on the analytic structure of the polynomial $G$ and is characterized by the so-called proper critical points. These are the stationary points at which $G$ does not
vanish and are, thus, defined as the solutions of 
\begin{equation}
\label{eq:critical-point-condition}
\frac{\partial G}{\partial \alpha_i} = 0 \  (i = 1, \ldots, m) \qquad
\mbox{and} \qquad G \neq 0.
\end{equation}
Proper critical points may or may not exist for a given $G$ polynomial. If there is no critical point, this implies that the corner integral of the sector is reducible, so that this sector does not contribute a master integral. On the other hand, if critical points do exist, the corner integral is irreducible with respect to IBP. This condition of irreducibility was first shown in \cite{Baikov:2005nv}.

A proper critical point may have a {\it multiplicity} which is measured by a topological invariant, the so-called Milnor number at the critical point.
The number of master integrals is the sum over the Milnor numbers $M_i$ of all proper critical points,
\begin{equation}
\# ~ \textrm{of master integrals} = \sum_{i \in \textrm{proper critical points}} M_i \,, 
\end{equation}
provided that the proper critical points are isolated. We will explain the meaning of {\it isolated critical point} shortly. In the case of non-isolated critical points, more work is required \cite{Lee:2013hzt}, see also an explicit example in section \ref{subset:mint-examples} below. 

Thus,  the problem of counting master integrals is reduced to the one of computing the critical points of a system of nonlinear polynomial equations given by
equation (\ref{eq:critical-point-condition}), which is usually highly nontrivial to solve. Fortunately, one can apply powerful algebraic approaches to make this job simple, based on the so-called Gr\"obner basis technique. Readers who are not familiar with the Gr\"obner basis and related concepts may consult \cite{cox2007ideals} for a pedagogical introduction. Here, only the basic procedure is outlined.

The solution space of a system of polynomial equations is called affine variety associated with an ideal determined by the equations. For the system of equations (\ref{eq:critical-point-condition}), the ideal can be defined as
\begin{equation}
\label{eq:ideal}
I = \left\langle \frac{\partial G}{\partial \alpha_1} \,, \, \ldots \,, \frac{\partial G}{\partial \alpha_m} \,, \, \alpha_0 G - 1 \right\rangle\, ,
\end{equation}
where, in the last term, one introduces an additional parameter $\alpha_0$ which forces the polynomial $G$ to be non-vanishing at the critical point. The solutions of the affine variety can be obtained by computing the Gr\"obner basis of 
equation (\ref{eq:ideal}),
\begin{equation}
{\rm gb}(I) = \left\{ g_1, g_2, \ldots, g_k \right\}\, .
\end{equation}
Once the Gr\"obner basis is obtained, it becomes relatively trivial to find the solutions. In our problem, we only concern ourselves with the number of solutions. This, in practice, can be conveniently obtained by counting the number of irreducible monomials in the Gr\"obner basis.\footnote{More precisely, one needs to compute the reduced Gr\"obner basis, which is unique for a given monomial ordering, see e.g.\ \cite{cox2007ideals}. The number of irreducible monomials also takes the multiplicity of critical points (i.e.\ the Milnor number) into account.} 

The above-described procedure is implemented in {\tt Mint} \cite{Lee:2013hzt} and applies straightforwardly to many simple cases; for example, applying it to the three-loop form factor, we obtain the result summarized in Table \ref{table:master-mint-3loop}. The numbers of master integrals are consistent with the reduction of {\tt LiteRed}. 

\begin{table}[ht]

\caption{Master integral counting for the six three-loop form factor topologies given in Figure~5 of \cite{Boels:2012ew}. }
\centering
\begin{tabular}{c | c c c c c c c c c c c c c c c c c } 
\hline %inserts double horizontal lines
\hline 
topology & (1) & (2) & (3) & (4) & (5)  & (6)  \\ [0.5ex] 
\hline 
\# MIs via {\tt Mint} & 9 & 10 & 10 & 10 & 14 & 10  \\ 
\hline 
\end{tabular}
\label{table:master-mint-3loop}
\end{table}  

For the four-loop form factor, however, one encounters two problems when trying to apply the {\tt Mint} package. The first problem is that the set of critical points can form an affine variety of dimension $\geq 1$. This corresponds to the non-isolated critical-point case. {\tt Mint} cannot deal with such cases automatically in its present version, but denotes them as {\it Indeterminate}. The number of such cases for the integral topologies at hand are summarized in Table \ref{table:indeterminate}. Although the absolute number of these cases is low, they tend to occur in the more complicated sectors. They require some further work as described in \cite{Lee:2013hzt}. We employed a similar procedure and present an explicit example in subsection \ref{subset:mint-examples}.

\begin{table}[ht]

\caption{{\it Indeterminate} sectors from {\tt Mint} for the 34 topologies given in Figures 8 and 9 of \cite{Boels:2012ew}.}
\centering
\begin{tabular}{c | c c c c c c c c c c c c c c c c c } 
\hline %inserts double horizontal lines
\hline 
topology & (14) & (15) & (16) & (17) & (18)  & (20) & (22) & (23) & (24)   \\ [0.5ex] 
\hline 
\# Indeterminate & 1 & 1 & 1 & 1 & 1 & 2 & 1 & 1 & 1  \\ 
\hline 
\hline 
topology & (25) & (26) & (27) & (28) & (29) & (30) & (32) & (33) & (34)  \\ [0.5ex] 
\hline 
\# Indeterminate & 1 & 3 & 1 & 1 & 3 & 1 & 1 & 1 & 1  \\ 
\hline 
\end{tabular}
\label{table:indeterminate}
\end{table} 

Another problem is that, in quite a lot of cases, {\tt Mint} gets stuck at the step of {\it Constructing Gr\"obner basis \dots}. This is due to the complexity of four-loop integrals for which {\tt Mathematica} cannot finish the computation of the corresponding Gr\"obner basis. There are many known other packages which can compute a Gr\"obner basis more efficiently. We applied {\tt Macaulay2} \cite{M2}, which can be conveniently used in {\tt Mathematica} through {\tt mathematicaM2.m} \cite{mathematicaM2}, and we employed {\tt Singular} \cite{DGPS} in
several most complicated cases. 

Based on {\tt Mint} and with some extra effort to solve the above two problems, we obtained the numbers of master integrals for the $34$ different topologies as summarized in Table \ref{table:master-mint}.

\begin{table}[ht]
\caption{Master integral statistics: {\tt Mint} results for the 34 topologies given in Figures 8 and 9 of \cite{Boels:2012ew}.}
\centering
\begin{tabular}{c | c c c c c c c c c c c c c c c c c       } 
\hline %inserts double horizontal lines
\hline 
topology & (1) & (2) & (3) & (4) & (5) & (6) & (7) & (8) & (9) & (10) & (11) & (12)  \\ [0.5ex]
\hline 
\# MIs & 28 & 35 & 38 & 34 & 39 & 48 & 71 & 52 & 55 & 69 & 52 & 95  \\ 
\hline 
\hline 
topology & (13) & (14) & (15) & (16) & (17) & (18) & (19) & (20) & (21) & (22) & (23) & (24)   \\ [0.5ex] 
\hline 
\# MIs  & 76 & 87 & 95 & 93 & 106 & 84 & 105 & 89 & 45 & 66 & 41 & 75  \\ 
\hline 
\hline 
topology & (25) & (26) & (27) & (28) & (29) & (30) & (31) & (32) & (33) & (34) \\  [0.5ex]
\hline 
\# MIs & 55 & 78 & 92 & 69 & 93 & 84 & 35  & 33 & 31 & 39 \\ 
\hline 
\end{tabular}
\label{table:master-mint}
\end{table} 

Given the number of master integrals, {\tt Mint} can also generate a set of master integrals. In most cases, a sector contains only one master integral, in which case the corner integral is chosen as the master integral. If there are more than one master integrals in a sector, {\tt Mint} suggests integrals with higher powers of propagators (while {\tt Reduze} typically chooses integrals with numerators). However, the choice of a double-propagator integral is rather heuristic; therefore, to find out whether it is a genuine master integral requires further checks. In our case, we could check them by using {\tt Reduze}, as will be discussed shortly.

To obtain the master integrals for the four-loop form factor, we still need to combine the master integrals of Table \ref{table:master-mint} together. Obviously, same master integrals may appear in different topologies of Table \ref{table:master-mint}. This may be detected by comparing their graph topology: if two graphs are isomorphic to each other, they are the same integral. %\footnote{We comment on two technical points. Master integrals are given in terms of propagators. It is usually subtle to obtain a graph representation from a given set of propagators. For us, this is straightforward, since we can start from the $34$ mother topologies and obtain the corresponding graphs for master integrals by removing propagators. Another technical point is about checking graph isomorphism via {\tt Mathematica}. The present version of {\tt Mathematica} cannot apply {\tt IsomorphicGraphQ} directly to mixed graphs. A simple trick to solve this problem is to add a vertex to each edge and then apply {\tt IsomorphicGraphQ}.} 
After this step, we find that there are only $280$ non-isomorphic master integrals: $244$ with only simple propagators plus $36$ containing a double propagator.\footnote{Note that we combined all 34 topologies including six zero-color-factor topologies: (8), (11), (15), (16), (18) and (20). If we exclude these six topologies, the number is reduced to $267$, which is enough for the ${\cal N}=4$ SYM form factor. But for QCD with fundamental quarks, one likely needs all $280$ master integrals.}

Here, one should be cautious that there could be relations between integrals which are apparently not due to graph isomorphism, see e.g.\ \cite{vonManteuffel:2012np,Pak:2011xt} for some explicit examples. Such relations can be found by considering Feynman parameter representations of integrals and checking if the two representations are related to each other by a permutation of Feynman parameters. A fast algorithm for such a check was proposed in \cite{Pak:2011xt}. Alternatively, one can also consider the so-called {\it graph matroid} technique to detect such hidden relations between different integrals, as implemented in {\tt Reduze} \cite{vonManteuffel:2012np}. For us, there is an easier way to detect such relations, namely applying {\tt Reduze} to reduce the basis integrals obtained by {\tt Mint}. We checked that all $244$ corner integrals correspond to basis integrals as determined by {\tt Reduze}. Thus, they are indeed master integrals.

As mentioned above, the choice of master integrals with double propagators by {\tt Mint} requires further checks. We applied {\tt Reduze} to the $32$ double-propagators integrals which do not appear in the twelve-propagator sector of topology $26$ and found that four of the suggested master integrals are actually reducible.\footnote{Note that {\tt Reduze} tends to choose integrals with numerators rather than double propagators. So, {\it irreducible} means that, in the reduction result of {\tt Reduze}, there is an integral with the same propagators containing a numerator.} This does not change the number of {\tt Mint} master integrals, but simply means that a different integral from this sector must be chosen as a master integral. By changing the position of the double propagator, we indeed found irreducible integrals. The twelve-propagator sector of integral topology $26$ is analyzed further below.  Thus, we obtained the final list of $280$ genuine master integrals. 

We would like to emphasize that the counting of master integrals based on {\tt Mint} only relies on the topologies of given integrals and applies to arbitrary numerators; therefore, it applies to any theory, including QCD. We remind that the counting of {\tt Reduze} given in Table \ref{table:counting_Reduze} concerns particular master integrals appearing in the reduction of the ${\cal N}=4$ SYM form factor result.

\subsection*{An interesting mismatch between {\tt Reduze} and {\tt Mint}}

Some caution is needed when assessing the output of {\tt Mint}: this package is known to under-report the number of master integrals in very rare cases. 
 
To check this, we collected all master integrals appearing in the IBP reduction
by {\tt Reduze}, with $t=12$ and $s=2$ and with $t=13$ and $s=1$. It turns out that {\tt Reduze} yields more master integrals. Many of them consist of twelve propagators and one to several numerators. They are at the edge of the reduction setting, and it is very likely that they may be reduced by including more IBP identities. These will not be discussed further here, details are available on request. What is unexpected is that there are nine corner integrals with only ten or less propagators, for example
\begin{equation}
\textrm{INT}[2, \{0, 0, 1, 1, 1, 1, 1, 1, 0, 0, 0, 1, 0, 0, 1, 0, 0, 0\}] \,,
\end{equation}
with only eight propagators. {\tt Mint} shows that they contain no proper critical points, while {\tt Reduze} and {\tt FIRE5} take them as master integrals.
Due to the low number of propagators, most related IBP identities should already be taken into account by {\tt Reduze}. Therefore, it is likely that these integrals are truly master integrals.\footnote{Some other examples of two-loop integrals showing mismatch between {\tt Mint} and IBP reduction were also found by V. Smirnov \cite{smirnov-private}.} This would mean that the counting method of {\tt Mint} requires some modification, although it is correct in the vast majority of cases. We cannot rule out, however, that there are integral relations which render this set of nine corner integrals reducible. It would be very interesting to understand this issue and see which is right.

In addition, there is one more master integral in the {\tt Reduze} basis, which appears to be an artifact of the way we solved the permutation symmetry consistency condition. 

\subsection{Two worked-out examples}
\label{subset:mint-examples}

Here, we give further details by considering two examples which cannot be computed directly by {\tt Mint}. 

Let us first consider an example corresponding to an indeterminate case: a subsector of topology $26$ with eleven propagators,
\begin{equation}
\textrm{INT}[26, \{1, 1, 1, 0, 1, 1, 1, 1, 1, 1, 1, 1, 0, 0, 0, 0, 0, 0\}] \,.
\end{equation} 
From $G=U+F$, which are polynomials of $\alpha_i$ $(i=1,\ldots,11)$ and of homogeneity five, one can construct the ideal $I$ defined in equation (\ref{eq:ideal}). The corresponding Gr\"obner basis cannot be computed by {\tt Mathematica}, but can be calculated using {\tt Macaulay2} \cite{M2}. The quotient space of the resulting ideal ${\cal I} = {\rm GB}(I)$ turns out to be infinite dimensional.\footnote{The dimensionality of the quotient space is given by the number of irreducible monomials, which can be directly counted from the reduced Gr\"obner basis as implemented in {\tt Mint} \cite{Lee:2013hzt}.} One then computes the primary decomposition of ${\cal I}$ (e.g.\ using {\tt Macaulay2}), which gives
\begin{equation}
{\cal I} = {\cal I}_1 \cap {\cal I}_2 \,,
\end{equation}
where
\begin{align}
{\cal I}_1 = & \langle 186624 \alpha _0-3125,\alpha _1,5 \alpha _2+12,5 \alpha _3-12,5 \alpha _4+12,5 \alpha _5-12,\alpha _6,5 \alpha _7+12, \nonumber\\
&  5 \alpha _8+12,5\alpha _9-12,5 \alpha _{10}+12,5 \alpha _{11}+12 \rangle   \,, \nonumber \\
{\cal I}_2 = & \langle 729 \alpha _0-16,2 \alpha _1+2 \alpha _8+3,\alpha _2+2 \alpha _{11}+6,\alpha _3-3,2 \alpha _4-2 \alpha _8+3,\alpha _5-3, \nonumber\\
& 2 \alpha _6+2  \alpha _{11}+3,\alpha _7+2 \alpha _8+6,2 \alpha _{11} \alpha _8+3 \alpha _8+3 \alpha _{11}+9,2 \alpha _9-3,2 \alpha _{10}-2 \alpha
   _{11}+3 \rangle   \,. \nonumber
\end{align}
The quotient space of the first ideal is one-dimensional, which is obvious, since there is only one solution of the corresponding polynomial system. On the other hand, the quotient space of the second ideal is infinite dimensional. This can be treated as the example discussed in \cite{Lee:2013hzt}. Rather than considering the full ideal ${\cal I}_2$, one can simply consider the variety determined by the equation ${\tilde G}(\alpha_8, \alpha_{11})=0$, where 
\begin{equation}
{\tilde G}(\alpha_8, \alpha_{11}) = 2 \alpha _{11} \alpha _8+3 \alpha _8+3 \alpha _{11}+9 
\end{equation}
is the only quadratic element in ${\cal I}_2$. Then, one computes the dimensionality of the quotient space of the ideal
\begin{equation}
{\tilde I} =  \left\langle {\partial {\tilde G} / \partial \alpha_8}  \,, {\partial {\tilde G} / \partial \alpha_{11}} \,, \, \alpha_0 {\tilde G} - 1 \right\rangle \, ,
\end{equation}
which turns out to be one. Altogether, one finds that the number of bases is two.

As another illustrative example, we consider the sector of topology $26$ with full twelve propagators. In this case, the computation of the Gr\"obner basis turns out to be hard even for {\tt Macaulay2}. In this case, we apply {\tt Singular}, which solves the problem in less than a minute using the method {\it slimgb}. The quotient space of the resulting Gr\"obner basis ${\cal I}={\rm GB}(I)$ is seven-dimensional. One still needs to consider the symmetry which may reduce the number of independent master integrals. In order to make the symmetry obvious, one can compute the primary decomposition of ${\cal I}$, which gives
\begin{equation}
{\cal I} = {\cal I}_1 \cap {\cal I}_2 \cap {\cal I}_3 \cap {\cal I}_4 \cap {\cal I}_5 \cap {\cal I}_6 \,,
\end{equation}
where
\begin{align}
{\cal I}_1 = & \langle 729 \alpha _0-16,4 \alpha _1+3,\alpha _2-3,\alpha _3-3,\alpha _4,4 \alpha _5+9,\alpha _6-3,\alpha _7-3,
2 \alpha _8+9, 4 \alpha _9+3,  \nonumber\\
&  2 \alpha _{10}-3,\alpha _{11}+6,2 \alpha _{12}+9 \rangle   \,, \nonumber \\
{\cal I}_2 = & \langle 729 \alpha _0-16,\alpha _1-3,2 \alpha _2+9,\alpha _3-3,\alpha _4,\alpha _5+6,\alpha _6-3,4 \alpha _7+3,\alpha _8-3, 2 \alpha _9+9, \nonumber\\
&  2 \alpha_{10}-3,4 \alpha _{11}+9,4 \alpha _{12}+3 \rangle   \,, \nonumber \\
{\cal I}_3 = & \langle 729 \alpha _0-16,2 \alpha _1+9,\alpha _2-3,2 \alpha _3+9,2 \alpha _4-3,\alpha _5+6,4 \alpha _6+3,\alpha _7-3,4 \alpha _8+3,  \nonumber\\
&  \alpha_9-3,\alpha _{10},4 \alpha _{11}+9,\alpha _{12}-3 \rangle   \,, \nonumber \\
{\cal I}_4 = & \langle 729 \alpha _0-16,\alpha _1-3,4 \alpha _2+3,4 \alpha _3+3,2 \alpha _4-3,4 \alpha _5+9,2 \alpha _6+9,2 \alpha _7+9,\alpha_8-3,  \nonumber\\
&  \alpha_9-3,\alpha _{10},\alpha _{11}+6,\alpha _{12}-3 \rangle   \,, \nonumber \\
{\cal I}_5 = & \langle 256 \alpha _0+1,\alpha _1+2,\alpha _2+2,\alpha _3+2,\alpha _4-2,\alpha _5+6,\alpha _6+2,\alpha _7+2,\alpha _8+2,\alpha _9+2, \nonumber\\
&  \alpha_{10}-2,\alpha _{11}+6,\alpha _{12}+2 \rangle   \,, \nonumber \\
{\cal I}_6 = & \langle 65536 \alpha _0-3125,5 \alpha _1+5 \alpha _{12}-8,5 \alpha _2+5 \alpha _{12}-8,\alpha _3-\alpha _{12},20 \alpha _4+5 \alpha_{12}-16, \nonumber\\
&  5 \alpha _5+16, \alpha _6-\alpha _{12},5 \alpha _7+5 \alpha _{12}-8,5 \alpha _8+5 \alpha _{12}-8,\alpha _9-\alpha _{12}, \nonumber\\
&  20 \alpha _{10}+5 \alpha_{12}-16,5 \alpha _{11}+16,25 \alpha _{12}^2-40 \alpha _{12}+128\rangle   \,. \nonumber
\end{align}
The quotient space of the last ideal is two-dimensional, while the others are all one-dimensional. So, naive counting would give seven master integrals. 
Topology $26$, however, has permutation symmetry, see equation (\ref{eq:symmetry-top26}). It is not hard to see that the first four ideals are all related to each other by this symmetry; therefore, they contribute only one independent master integral. In total, one finds that there are four independent master integrals in this sector. For this integral topology, {\tt Reduze} picks four twelve-propagator integrals as master integrals: the corner integral, two linear-numerator integrals and one with a quadratic numerator. Hence, the number of master integrals matches between {\tt Mint} and {\tt Reduze}.

\section{Towards numerical integration}\label{sec:towardsnumerics}

Given a set of master integrals, the next step is to perform the integration. A significant part of the master integrals obtained in the previous sections
are very complicated and seem to be on or slightly over the very edge of what current integration methods can achieve. There are various methods to perform numerical integration in the literature, see \cite{Smirnov:2012gma} for an overview. 

It should be said that, in principle, Mellin--Barnes (MB) representations, which can be used very efficiently in solving planar topology integrals \cite{Czakon:2005rk, Smirnov:2009up, Gluza:2007rt, Gluza:2010rn}, may be the optimal weapon of choice for four-loop integrals, if the difficulties in obtaining efficient MB representations for non-planar integrals without hidden singularities can be resolved. See \cite{Smirnov:2012gma}, page 124, for a brief description of the problem. Progress in obtaining up to and including some three-loop non-planar representations in an automatic way was reported in \cite{Blumlein:2014maa}. This may be used, at least in principle, to obtain MB representations for those integral topologies containing either a bubble or a triangle with two on-shell legs. 

We mostly explored numerical integration using sector decomposition \cite{Binoth:2000ps}. In general, this is a method to turn a highly divergent integral into a sum over finite integrals, by resolving the singularities of the Feynman parameter integral. These finite integrals can then be integrated, for instance, using the {\tt CUBA} library \cite{Hahn:2004fe}. We are aware of three different implementations of this idea in public codes: {\tt sector\_decomposition} \cite{Bogner:2007cr}, {\tt SecDec} \cite{Carter:2010hi, Borowka:2012yc, Borowka:2015mxa} and {\tt FIESTA} \cite{Smirnov:2008py, Smirnov:2009pb, Smirnov:2013eza}. Only the latter program in its latest versions (3.x) was able to handle some of the most difficult integrals of the problem at hand.

\subsection{Relevant implementation details}
\label{subsec:numerical-details}

There are various strategies for resolving the singularities of a given Feynman integral. {\tt FIESTA} offers a choice of these. Most relevant here are strategies {\it S} \cite{Smirnov:2008py} and {\it X} \cite{Binoth:2000ps}. Strategy {\it KU} \cite{Kaneko:2009qx}, which, in principle, is highly efficient in terms of
sectors, did not resolve any of the sectors of the most complicated integrals (the corner integral of topology 25) in either of its three variants and was not pursued further. Strategy {\it X}, if it terminates, tends to yield better results in terms of the number of sectors than strategy {\it S}.

{\tt FIESTA} splits the Feynman parameter integral in equation (\ref{eq:defalphaparam}) into different {\it primary sectors}. In each primary sector, one particular Feynman parameter is smaller than the others. If the integrand is invariant under a permutation of the $\alpha$ parameters, then these primary sectors may be related. A particular permutation symmetry of the $\alpha$ parameters descends from the parent graph. This symmetry may then be used to reduce the computational workload. 
%Let us consider, for instance, the scalar integral in topology $26$, with the propagators as listed in equation \ref{eq:propstop26}. The two independent permutation symmetries of the graph correspond to the following cycles:
%\begin{equation}
%\label{eq:symmetry-top26II}
%\begin{array}{c} 
%\{ \{ 1, 7 \}, \{ 2, 8 \}, \{ 3, 6 \}, \{ 5, 11 \}, \{ 9, 12 \} \} \,, \\
% \{ \{ 1, 2 \}, \{ 3, 9 \}, \{ 4, 10 \}, \{ 6, 12 \}, \{ 7, 8 \} \} \,.
%\end{array}
%\end{equation}
%Together with the trivial permutation and their product, these form a four-element representation of the permutation group. A given propagator can be mapped to other propagators in the set, but only if they are in the orbit of the given permutation. The orbits of the above permutation group are
%\begin{equation}
%\{\{1,2,7,8\},\{3,6,9,12\},\{4,10\},\{5,11\}\} \, .
% \end{equation}
Let us recall the symmetry property of the scalar integral of topology $26$ given in equation (\ref{eq:symmetry-top26}). From its orbits of the permutation group in equation (\ref{eq:orbits-group-26}), 
\begin{equation}\label{eq:top26orbits}
\{\{1,2,7,8\},\{3,6,9,12\},\{4,10\},\{5,11\}\} \, .
 \end{equation}
we can see that, 
instead of computing primary sectors $1$, $2$, $7$ and $8$ separately, one better just computes one of these, as the others will give the same numerical answer. In practice, it can matter which coefficient is picked, since the structure of the integrand for the $\alpha$ parameters is different; this can result in different numbers of sectors. 

The integration time varies; for integrals with twelve propagators, the preparation time alone is typically measured in days to weeks. By contrast, all integrals with less than eleven propagators are highly tractable using {\tt FIESTA}, typically requiring less than a day.

\subsection*{Quadratic-numerator integral in topology 26}

The quadratic-numerator integral in topology $26$ is naively expected to be one of the hardest integrals in the full set. It can be integrated using {\tt FIESTA}. Integral topology $26$ has two permutation symmetries, see equation (\ref{eq:premium}). To make use of this, one needs to pick a symmetric quadratic numerator. That is, one needs a numerator which is mapped to itself under both symmetries and which is  related to the quadratic-numerator master integral in the output of {\tt Reduze} by IBP reduction. Moreover, to facilitate the computation of the double derivative in the $\alpha$ parametrization, it is advantageous to pick a numerator that is linear in the loop momenta. It is easy to check that 
\begin{equation}
\left( l_3 \cdot (p_1 - p_2)\right)^2 
\end{equation}
transforms into itself under the symmetries in equation (\ref{eq:premium}). We checked that this is related to the quadratic master integral from the explicit IBP reduction. With the symmetries in place, one only needs to compute four out of twelve primary-sector coefficients in {\tt FIESTA}. We picked the coefficients $\{4,0,4,2,2,0,0,0,0,0,0,0\}$ for the order of propagators in equation (\ref{eq:propstop26}), basically choosing the lowest from the orbits in equation (\ref{eq:top26orbits}).

The resulting integral can be integrated with some numerical effort, giving the result
\begin{align}
\textrm{INT}_{26}  [ ( l_3 \cdot (p_1 - p_2)) ^2 ] = & (-0.032986 \pm 2 \cdot 10^{-7}) \, \epsilon^{-8} +
(0.0694456  \pm 1.01 \cdot 10^{-5}) \, \epsilon^{-7}  \nonumber\\  & +  (1.3506  \pm 0.0002)  \,\epsilon^{-6} + (-2.68804 \pm 0.00317) \, \epsilon^{-5} \nonumber\\
& +  (-6.23707 \pm 0.04013)  \, \epsilon^{-4} + (12.6763 \pm 2.0782) \, \epsilon^{-3} \nonumber\\ 
&+  (1234.49 \pm 32.97)  \,\epsilon^{-2} + {\cal O}(\epsilon^{-1}) \,.
\end{align}
Preparing for the numerical integration takes about five days. The integration time itself strongly depends on the integration parameters in the {\tt Vegas} integration algorithm of the {\tt CUBA} library.\footnote{A brief exploration of the other algorithms in the {\tt CUBA} library quickly shows that {\tt Vegas} is the optimal choice here, as for both time and accuracy.} We checked by running with different numbers of evaluation points that the first five coefficients are numerically stable under variations of the relevant parameters, especially {\tt maxeval}. In other words, it is likely that their errors are in the reported ranges. 

%\subsection{Missing results for the planar form factor}
%
%We aimed at computing the planar form factor first, as all important cross-checks can be made there. For most integrals, several terms of the $\epsilon$ expansion were obtained. However, three integrals have resisted computation so far. These are the twelve-propagator corner integrals of topologies $25$ and $30$, as well as an additional twelve-propagator master integral from topology $30$. These integrals can be used as a simple test to check if any attempt to compute the four-loop form factor integrals can be successful. 

%%% [ this may be more fitted in the discussion ]

\section{Summary and discussion}

%This article represents a major step toward the computation of four-loop Sudakov form factor as well as the non-planar cusp anomalous dimension. 
In this paper, we obtained an explicit integral reduction for the four-loop Sudakov form factor in $\mathcal{N}=4$ SYM based on the integrand result in \cite{Boels:2012ew} and by exploiting the {\tt Reduze} package \cite{vonManteuffel:2012np}. This reduces the challenging four-loop problem, in particular the computation of the unknown non-planar cusp anomalous dimension, down to the evaluation of an explicit set of given master integrals. 

We also obtained master integrals by using the algebraic method introduced in \cite{Lee:2013hzt}. This provides a first non-trivial implementation of the method to a complete observable. Limits of the {\tt Mint} package and their resolutions are explained. The resulting integrals are, with some qualifications, the master integrals of quite generic form factors. For instance, they are expected to cover all integrals which appear in the QCD result. By comparing with the results of IBP reduction, we found interesting mismatches, namely, existing masters integrals from IBP reduction are taken to be reducible by the {\tt Mint} package. It would be important to understand the reason of this discrepancy.
%Obtaining an explicit IBP reduction for the case of QCD seems to be out of reach for currently available technology, as it involves up to eight powers of numerators. 

Based on the reduction result, we furthermore observed surprising cancellations in the form factor. First, the form factor integrand obtained in \cite{Boels:2012ew} contains a free parameter. While it was unclear whether this parameter could be uniquely fixed by further physical constraints, we showed that the terms depending on this parameter all magically cancel upon using IBP identities. Therefore, this parameter is truly free and gives the first example of a color-kinematic-dual integrand with such a free parameter.  %These two cases are among the classes of zero color factors. 
Furthermore, two integrals, corresponding to topologies $18$ and $20$, turn out to be exactly zero, which is not obvious at all at the integrand level. It would be very interesting to have a physical understanding of these intriguing cancellations.

What remains to be done is to compute the master integrals. However, for many of them, this is currently just out of reach for publicly available (implementations of) integration methods. There is an important analogue to the three-loop case, which we would like to draw attention to. In that case, a reduction to master integrals was obtained in \cite{Gehrmann:2006wg}. Their integration was first
performed almost three years later in \cite{Baikov:2009bg}, using partly numerical methods. An analytic result was then reported in \cite{Lee:2010cga}, with an important cross-check in \cite{Gehrmann:2010ue}. Currently, the three-loop form factor integrals often appear as an example application of new integration and organization methods, e.g. in \cite{Henn:2013nsa,vonManteuffel:2014qoa}. 

A new feature in the four-loop case is the appearance of several master integrals beyond corner integrals. These can be taken to be either doubled-up propagators or single numerators. The latter can, in principle, be traded for higher-dimensional scalar integrals using the Feynman parametrization in equation (\ref{eq:defalphaparam}). The problem is that these scalar integrals feature four {\it dots}: there is a sum over integrals with four doubled-up propagators. Without an IBP reduction of these integrals down to the basis, this is an unpractical result. An explicit IBP reduction with up to four dots would enable the use of several methods. For instance, one could use dimensional recurrences \cite{Tarasov:1996br} to either compute the integrals numerically \cite{Lee:2009dh} or to express the form factor in terms of a quasi-finite basis \cite{vonManteuffel:2014qoa}. An explicit quasi-finite integral reduction for the three-loop form factor has been announced \cite{vonmanteuffelradcor}. The needed IBP reductions at four loops seem very hard to obtain within the current version of {\tt Reduze} and simply out of reach for anything else on the public market. See \cite{vonManteuffel:2014ixa} for a proposal of an improved algorithm which is particularly suitable for single-scale reduction problems. 

There is a practical issue when calculating the form factor which concerns the choice of master integrals. There are several choices of master integrals one
could make as for both linear numerators and doubled propagators. Which one to choose should then be guided by {\it integrability}, i.e.\ by which of these choices they can actually be integrated. Moreover, as for numerical evaluation, there are issues of precision and speed to be considered. One would expect doubled-up propagators to give simpler Feynman parameter integrals, while linear numerators tend to give less divergent coefficients. A canonical example of the two choices are the master integrals of the double-box-integral topology. One can either choose a doubled-propagator basis element, as first done in \cite{Smirnov:1999wz}. Alternatively, one can use a linear-numerator integral, as in \cite{Anastasiou:2000kp}. The latter has a less divergent prefactor as a function of $\epsilon$. Which choice is the more practical must be decided on a case-by-case basis though: we observed cases where no difference in expansion order is present. 

There are various other avenues to be explored which may lead to a full computation of the cusp anomalous dimension through the form factor route. One would be to explore MB representations of four-loop non-planar integrals. An obstacle here is that naive implementations of this idea quickly lead to high-order MB representations containing an order of one hundred MB integrations. Any algorithm which could curb this number would offer a way forward. While numerical evaluation may have the precision issue, analytic methods would be more desirable. Linearly reducible integrals may be evaluated analytically by {\tt HypInt} \cite{Panzer:2014caa}. This works particularly efficiently for integrals of less singular behavior \cite{vonManteuffel:2014qoa}. An alternative route for computing the cusp anomalous dimension is to extend integration techniques based on differential equations \cite{Henn:2013pwa} to form factors by making one extra leg off-shell and then taking the massless limit. Controlling this limit after integration might also be an obstacle. Apart from that, IBP reduction, a necessary step in constructing the differential equation, would be more challenging for deformed integrals. See \cite{Henn:2013nsa} for an application to the three-loop case.

\subsection*{Acknowledgements}

We would like to thank Oleg Tarasov for collaboration at the initial stages of this project. It is a pleasure to thank Roman Lee, Andreas von Manteuffel, Sven Moch, Volodya Smirnov, Alexander Smirnov and Yang Zhang for discussions. This work was supported in part by the German Science Foundation (DFG) through the Collaborative Research Center SFB~676 ``Particles, Strings and the Early Universe: the Structure of Matter and Space-Time" and a DFG grant in the framework of the SFB~647 ``Space -- Time -- Matter".

\appendix

\section{Three-loop form factor check}

The three-loop form factor was first computed in \cite{Baikov:2009bg}; we started from the expression obtained in \cite{Boels:2012ew}, which can be shown to be equivalent. The problem consists of six integrals with at most a single numerator. The integrals contain nine propagators, and three additional numerators have to be specified to obtain a definite basis. Five of these integrals have a single permutation symmetry. 

IBP reduction using {\tt Reduze} on a part of our local computing cluster running in parallel mode took about 1.5 days. The reduction yielded a set of master integrals, the most complicated of which indeed correspond to those obtained in \cite{Gehrmann:2006wg}. These integrals were then computed using {\tt FIESTA}, running for a few days with their standard precision setting. The result for the form factor is
\begin{multline}
F_{2,{\rm num}}^{(3)} = -\frac{0.166606}{\epsilon ^6}-\frac{0.00012}{\epsilon ^5}+\frac{0.00054}{\epsilon
   ^4}+\frac{1.10182}{\epsilon ^3}+\frac{0.93788}{\epsilon ^2}+O\left(\frac{1}{\epsilon}\right) \, ,
\end{multline}
which is to be compared to the known result, e.g.\ from \cite{Gehrmann:2011xn},
\begin{equation}
F_{2,{\rm exact}}^{(3)} = -\frac{1}{6  \,\epsilon ^6} + \frac{11 \zeta (3)}{12  \, \epsilon ^3}+\frac{247 \pi ^4}{25920
  \, \epsilon ^2}+O\left(\frac{1}{\epsilon}\right) \, .
 \end{equation}
The difference between the exact answer and the numerical approximation is
\begin{equation}
\Delta F_2^{(3)} = \frac{0.000060}{\epsilon^6}-\frac{0.00012}{\epsilon
   ^5}+\frac{0.00054}{\epsilon ^4}-\frac{0.00006}{\epsilon
   ^3}+\frac{0.00964}{\epsilon ^2} +O\left(\frac{1}{\epsilon}\right) \, .
\end{equation}
Note that the final coefficient is only off by about one percent. With more numerical effort, this could be brought down more. For the purposes of this article, it is important that the calculational pipeline works, at least in principle.

\section{Alternative choice of numerators}

The following choice of numerator factors for master integrals optimizes the power series expansions of the coefficients to some extent. This choice is guided by the following principles:

\begin{itemize}
\item choose the numerators linear in the loop momenta to minimize the complexity of the Feynman parameter integrals;
\item when expressed in the master integral basis preferred by {\tt Reduze}, the coefficient of the linear-numerator master integral comes with a negative power of $\epsilon$ if possible;
\item when expressed in the master integral basis preferred by {\tt Reduze}, the coefficient of the scalar version of the linear-numerator master integral comes with a negative power of $\epsilon$;
\item preserve permutation symmetry if possible. 
\end{itemize}

This leads to the following list, where the coordinates are in the basis of the integral topologies given in appendix~\ref{app:basis}:
\begin{align*}
 &  \textrm{INT}[30, \{1, 1, 1, 1, 1, 1, 1, 1, 1, 1, 1, 1, 0, 0, 0, 0, 0, 0\}, p_2 \cdot l_5] \,,\\
 &  \textrm{INT}[3, \{1, 0, 0, 1, 0, 1, 1, 0, 1, 1, 0, 1, 0, 1, 1, 1, 1, 1\},  p_1\cdot (l_3 - l_5)] \,,\\
 &  \textrm{INT}[28, \{1, 1, 1, 1, 1, 1, 1, 1, 1, 1, 1, 1, 0, 0, 0, 0, 0, 0\},  p_2\cdot l_5] \,,\\
 &  \textrm{INT}[26, \{1, 1, 1, 1, 1, 1, 1, 1, 1, 1, 1, 1, 0, 0, 0, 0, 0, 0\},  p_1 \cdot l_4] \,,\\
 &  \textrm{INT}[25, \{1, 1, 1, 1, 1, 1, 1, 1, 1, 1, 1, 1, 0, 0, 0, 0, 0, 0\},  (p_2 - p_1)\cdot l_3] \,,\\
 &  \textrm{INT}[23, \{1, 1, 1, 1, 1, 1, 1, 1, 1, 1, 1, 1, 0, 0, 0, 0, 0, 0\},  p_2\cdot l_5] \,,\\
 &  \textrm{INT}[20, \{1, 0, 1, 1, 1, 1, 1, 1, 1, 1, 1, 1, 0, 0, 1, 0, 0, 0\},  p_2\cdot (l_5 + l_6 - l_3)] \,,\\
 &  \textrm{INT}[17, \{1, 0, 1, 1, 1, 1, 1, 1, 1, 1, 1, 1, 0, 0, 0, 1, 0, 0\},   p_2\cdot (l_3 - l_5)] \,,\\
 &  \textrm{INT}[11, \{1, 0, 1, 1, 1, 1, 1, 1, 0, 1, 1, 1, 0, 0, 1, 0, 0, 1\},   p_1\cdot l_5] \,,\\
 &  \textrm{INT}[5, \{1, 0, 1, 1, 1, 1, 1, 1, 1, 1, 1, 1, 0, 0, 1, 0, 0, 0\},   (p_2 - p1)\cdot l_3] \,,\\
 &  \textrm{INT}[4, \{1, 0, 1, 1, 1, 1, 1, 1, 1, 1, 0, 1, 1, 1, 0, 0, 0, 0\}, p_1\cdot l_3] \,,\\
 &  \textrm{INT}[3, \{1, 1, 0, 1, 1, 1, 1, 0, 1, 1, 0, 1, 0, 1, 1, 0, 1, 0\}, p_2\cdot (l_4 - l_5)] \,,\\
 &  \textrm{INT}[2, \{0, 0, 1, 1, 1, 1, 1, 1, 1, 1, 1, 1, 0, 0, 1, 1, 0, 0\}, p_1\cdot l_3] \,.
\end{align*}
This choice leads to a significant reduction in the IBP computational workload, at the expense of introducing non-symmetric numerators. 

\section{Basis}
\label{app:basis}

This appendix contains the main basis used in the reduction. The numbering of the equations corresponds to the topologies in \cite{Boels:2012ew}. In each case, the twelve entries given in the first line parametrize the twelve propagators of the respective integral and the six entries in the second line the chosen numerators. Note that the numerators do not respect the symmetries of the diagram topology corresponding to the first twelve propagators, with the exception of topology 1. We have defined $q = p_1 + p_2$.

\abovedisplayskip=1pt
\belowdisplayskip=1pt
\begin{multline}
\{   l_6  , \, l_5  , \, l_4  , \, l_3  , \,  l_6-l_5   , \,  l_4-l_3   , \,  p_1-l_6  , \, 
    l_5-l_4   , \,  -l_6+q   , \,  -l_4+q   , \,  -l_3+q  , \, 
    -l_5+q   , \\ 
    l_3-l_5   , \,  l_3-l_6   , \,  l_4-l_6   , \,  l_3-p_2  , \, 
    l_4-p_2   , \,  l_5-p_2 \}  \,,
\end{multline}
\begin{multline}
\{   l_6  , \, l_5  , \, l_4  , \, l_3  , \,  l_5-l_4   , \,  l_6-l_5   , \,  p_1-l_6  , \, 
    l_3-l_4   , \,  -l_3+q   , \,  -l_6+q   , \,  -l_5+q  , \, 
    -l_3+l_4-l_5+q   , \\  
     l_3-l_5   , \,  l_3-l_6   , \,  l_4-l_6  , \, 
    l_4-p_1   , \,  l_5-p_2   , \,  l_3-p_2    \}  \,,
\end{multline}
\begin{multline}
\{   l_6  , \, l_5  , \, l_4  , \, l_3  , \,  l_4-l_3   , \,  l_4-l_5   , \,  l_6-l_5  , \, 
    p_1-l_6   , \,  -l_4+q   , \,  -l_6+q   , \,  -l_3+q  , \, 
    -l_4+l_5-l_6+q   ,  \\  
     l_5-p_1   , \,  l_3-p_2   , \,  l_4-p_2  , \, 
    l_5-p_2   , \,  l_3-l_5   , \,  l_3-l_6    \}  \,,
\end{multline}
\begin{multline}
\{   l_6  , \, l_5  , \, l_4  , \, l_3  , \,  p_1-l_6   , \,  l_4-l_3   , \,  l_5-l_6  , \, 
    l_5-l_4   , \,  -l_5+l_6+p_2   , \,  -l_5+q   , \,  -l_3+q  , \, 
    -l_4+q   , \\   
     l_3-l_5   , \,  l_3-l_6   , \,  l_4-l_6   , \,  l_3-p_2  , \, 
    l_4-p_2   , \,  l_6-p_2    \}  \,,
\end{multline}
\begin{multline}
\{   l_6  , \, l_5  , \, l_4  , \, l_3  , \,  l_3-l_4   , \,  l_5-l_4   , \,  p_1-l_6  , \, 
    l_5-l_6   , \,  -l_5+q   , \,  -l_5+l_6+p_2   , \,  -l_3+q  , \, 
    -l_3+l_4-l_5+q   , \\    l_3-p_2   , \,  l_3-l_6   , \,  l_4-l_6  , \, 
    l_4-p_1   , \,  l_4-p_2   , \,  l_6-p_2    \}  \,,
\end{multline}
\begin{multline}
\{   l_6  , \, l_5  , \, l_4  , \, l_3  , \,  p_1-l_5   , \,  l_4-l_5   , \,  l_4-l_3  , \, 
    p_2-l_6   , \,  -l_3+q   , \,  -l_4+q   , \,  -l_4-l_6+q  , \, 
    -l_5-l_6+q   , \\  
    l_3-l_5   , \,  l_3-l_6   , \,  l_4-l_6  , \, 
    l_5-l_6   , \,  l_3-p_2   , \,  l_4-p_2    \}  \,,
\end{multline}
\begin{multline}
\{   l_6  , \, l_5  , \, l_4  , \, l_3  , \,  p_1-l_5   , \,  l_4-l_5   , \,  l_4-l_3  , \, 
    p_2-l_6   , \,  -l_4+q   , \,  -l_4+l_5+l_6   , \,  -l_3+q  , \, 
    -l_5-l_6+q   , \\  
    l_3-l_5   , \,  l_3-l_6   , \,  l_5-l_6  , \, 
    l_3-p_2   , \,  l_4-p_2   , \,  l_5-p_2   \}  \,,
\end{multline}
\begin{multline}
\{   l_6  , \, l_5  , \, l_4  , \, l_3  , \,  p_2-l_6   , \,  l_4-l_5   , \,  l_4-l_3  , \, 
    p_1-l_5   , \,  -l_4+q   , \,  -l_3+q   , \,  -l_4-l_6+q  , \, 
    l_4-l_5+l_6-p_2   , \\    
    l_3-l_5   , \,  l_3-l_6   , \,  l_4-l_6   , \,  l_5-l_6  
   , \,  l_3-p_2   , \,  l_5-p_2    \}  \,,
\end{multline}
\begin{multline}
\{   l_6  , \, l_5  , \, l_4  , \, l_3  , \,  l_4-l_3   , \,  p_1-l_5   , \,  l_4-l_5  , \, 
    -l_5+q   , \,  -l_3+q   , \,  -l_4-l_6+q  , \, 
    l_5+l_6-q   , \,  -l_3-l_6+q   , \\    
    l_3-l_5   , \,  l_3-p_2  , \, 
    l_4-p_2   , \,  l_4-l_6   , \,  l_5-l_6   , \,  l_6-p_2    \}  \,,
\end{multline}
\begin{multline}
\{   l_6  , \, l_5  , \, l_4  , \, l_3  , \,  l_4-l_5   , \,  p_1-l_5   , \,  l_3-l_4  , \, 
    l_3-l_4+l_6   , \,  -l_3+q   , \,  -l_5+q  ,  \, 
    -l_3-l_6+q   , \,  l_3-l_4+l_5+l_6-q   , \\    l_5-l_6  , \, 
    p_2-l_6   , \,  l_3-l_5   , \,  l_4-p_2   , \,  l_3-p_2   , \,  l_4-l_6   \}  \,,
\end{multline}
\begin{multline}
\{   l_6  , \, l_5  , \, l_4  , \, l_3  , \,  l_4-l_5   , \,  p_1-l_5   , \,  l_4-l_3  , \, 
    -l_3+q   , \,  -l_5+q   , \,  l_5+l_6-q  , \,
    -l_3-l_6+q   , \,  l_3-l_4+l_5+l_6-q   , \\    l_3-l_5  , \, 
    l_3-p_2   , \,  l_4-p_1   , \,  l_4-p_2   , \,  l_5-l_6   , \,  l_6-p_2   \}  \,,
\end{multline}
\begin{multline}
\{   l_6  , \, l_5  , \, l_4  , \, l_3  , \,  p_2-l_6   , \,  p_1-l_5   , \,  l_3-l_4  , \, 
    l_5-l_4   , \,  -l_3+q   , \,  -l_4+l_5+l_6   , \,  -l_5-l_6+q  , \,
    -l_3+l_4-l_5-l_6+q   , \\   
    l_3-l_6   , \,  l_5-p_2   , \,  l_3-p_2  , \, 
    l_4-p_1   , \,  l_4-p_2   , \,  l_6-p_1    \}  \,,
\end{multline}
\begin{multline}
\{   l_6  , \, l_5  , \, l_4  , \, l_3  , \,  l_3-l_4   , \,  p_2-l_6   , \,  p_1-l_4  , \, 
    l_6-l_5   , \,  -l_3+q   , \,  -l_4-l_6+q  , \, 
    -l_3-l_5+q   , \,  -l_4-l_5+q   , \\    l_3-l_6   , \,  l_3-p_2  , \, 
    l_4-p_2   , \,  l_5-p_1   , \,  l_5-p_2   , \,  l_6-p_1    \}  \,,
\end{multline}
\begin{multline}
\{   l_6  , \, l_5  , \, l_4  , \, l_3  , \,  p_2-l_6   , \,  p_1-l_4   , \,  l_3-l_4  , \, 
    l_6-l_5   , \,  -l_3+l_4+l_5   , \,  -l_3+q   , \,  -l_4-l_5+q  , \, 
    -l_4-l_6+q   , \\    l_4-l_5   , \,  l_3-l_6   , \,  l_4-l_6  , \, 
    l_3-p_2   , \,  l_4-p_2   , \,  l_5-p_2    \}  \,, 
\end{multline}

\begin{multline}
\{   l_6  , \, l_5  , \, l_4  , \, l_3  , \,  p_2-l_6   , \,  p_1-l_4   , \,  l_3-l_4  , \, 
    l_5-l_6   , \,  -l_3+q   , \,  -l_4-l_5+l_6+p_1  , \, 
    -l_3-l_5+q   , \,  -l_4-l_5+q   , \\    l_3-l_6   , \,  l_3-p_2  , \, 
    l_4-p_2   , \,  l_5-p_1   , \,  l_5-p_2   , \,  l_6-p_1    \}  \,, 
\end{multline}
\begin{multline}
\{   l_6  , \, l_5  , \, l_4  , \, l_3  , \,  p_2-l_6   , \,  p_1-l_4   , \,  l_6-l_5  , \, 
    l_3-l_4   , \,  -l_3+q   , \,  l_3-l_4+l_5-l_6  , \, 
    -l_3-l_5+q   , \,  -l_4-l_6+q   , \\    l_3-l_6   , \,  l_3-p_2  , \, 
    l_4-l_5   , \,  l_4-p_2   , \,  l_5-p_1   , \,  l_6-p_1     \}  \,,
\end{multline}
\begin{multline}
\{   l_6  , \, l_5  , \, l_4  , \, l_3  , \,  p_2-l_6   , \,  l_5-l_4   , \,  p_1-l_5  , \, 
    l_4-l_3   , \,  -l_3+q   , \,  -l_5-l_6+q  , \, 
    -l_3+l_4-l_5+q   , \,  -l_3+l_4-l_5-l_6+q   , \\    
    l_3-p_2  
   , \,  l_4-l_6   , \,  l_5-p_2   , \,  l_4-p_1   , \,  l_4-p_2   , \,  l_6-p_1    \}  \,,
\end{multline}
\begin{multline}
\{   l_6  , \, l_5  , \, l_4  , \, l_3  , \,  p_1-l_5   , \,  p_2-l_6   , \,  l_3-l_4  , \, 
    l_5-l_4   , \,  l_3-l_4-l_6   , \,  -l_3+q  , \,
    -l_3+l_4-l_5+q   , \,  -l_3+l_4-l_5+l_6+p_1   , \\    l_3-p_2  , \, 
    l_3-l_5   , \,  l_4-l_6   , \,  l_4-p_2   , \,  l_5-p_2   , \,  l_6-p_1    \}  \,,
\end{multline}
\begin{multline}
\{   l_6  , \, l_5  , \, l_4  , \, l_3  , \,  p_2-l_6   , \,  p_1-l_5   , \,  l_4-l_5  , \, 
    l_3-l_4   , \,  -l_4+l_5+l_6   , \,  -l_3+q   , \,  -l_3+l_5+l_6  , \,
    -l_5-l_6+q   , \\    
    l_3-l_6   , \,  l_5-l_6   , \,  l_4-p_1  , \, 
    l_3-p_2   , \,  l_4-p_2   , \,  l_5-p_2     \}  \,,
\end{multline}
\begin{multline}
\{   l_6  , \, l_5  , \, l_4  , \, l_3  , \,  p_2-l_6   , \,  l_4-l_3   , \,  p_1-l_5  , \, 
    l_5-l_4   , \,  -l_3+l_5+l_6   , \,  -l_3+q   , \,  -l_3+l_4+l_6  , \,
    -l_5-l_6+q   , \\   
    l_4-l_6   , \,  l_5-l_6   , \,  l_4-p_1  , \, 
    l_3-p_2   , \,  l_4-p_2   , \,  l_5-p_2     \}  \,,
\end{multline}
\begin{multline}
\{   l_6  , \, l_5  , \, l_4  , \, l_3  , \,  p_1-l_4   , \,  l_3-l_4   , \,  p_2-l_6  , \, 
    l_5+l_6   , \,  -l_4-l_5+p_1   , \,  -l_3+q   , \,  -l_3+l_6+p_1  , \,
    -l_3-l_5+p_1   , \\    
    l_3-p_2   , \,  l_4-p_2   , \,  l_5-p_2   , \,  l_6-p_1  , \, 
    l_4-l_6   , \,  l_3-l_5     \}  \,,
\end{multline}
\begin{multline}
\{   l_6  , \, l_5  , \, l_4  , \, l_3  , \,  l_3-l_4   , \,  l_5+l_6   , \,  p_2-l_6  , \, 
    p_1-l_4   , \,  -l_3+l_6+p_1   , \,  -l_3+q   , \,  -l_4-l_5+p_1  , \,
    -l_3+l_4+l_5+l_6   , \\    
    l_3-l_5   , \,  l_3-p_2   , \,  l_4-p_2  , \, 
    l_5-p_1   , \,  l_5-p_2   , \,  l_6-p_1     \}  \,,
\end{multline}
\begin{multline}
\{   l_6  , \, l_5  , \, l_4  , \, l_3  , \,  l_6-p_2   , \,  p_1-l_4   , \,  l_6-l_5  , \, 
    l_3-l_4   , \,  -l_3+q   , \,  l_3-l_4-l_5   , \,  -l_3+l_6+p_1  , \,
    -l_3+l_5+p_1   , \\  
    l_3-p_2   , \,  l_4-p_2   , \,  l_5-p_1   , \,  l_6-p_1  , \, 
    l_5-p_2   , \,  l_4-l_6     \}  \,,
\end{multline}
\begin{multline}
\{   l_6  , \, l_5  , \, l_4  , \, l_3  , \,  p_2-l_6   , \,  l_5+l_6   , \,  l_3-l_4  , \, 
    p_1-l_4   , \,  -l_3-l_5+p_1   , \,  -l_4-l_5+p_1   , \,  -l_3+q  , \,
    -l_3-l_5-l_6+q   , \\    
    l_3-p_2   , \,  l_4-l_6   , \,  l_4-p_2  , \, 
    l_5-p_1   , \,  l_5-p_2   , \,  l_6-p_1     \}  \,,
\end{multline}
\begin{multline}
\{   l_6  , \, l_5  , \, l_4  , \, l_3  , \,  p_1-l_5   , \,  l_5-l_4   , \,  p_2-l_6  , \, 
    l_3-l_4   , \,  -l_4+l_5+l_6   , \,  -l_3+q   , \,  -l_3+l_4-l_5+p_1  , \,
    -l_3+l_4-l_5-l_6+q   , \\    
    l_3-l_5   , \,  l_3-l_6   , \,  l_5-l_6  , \, 
    l_4-p_1   , \,  l_4-p_2   , \,  l_5-p_2     \}  \,,
\end{multline}
\begin{multline}
\{   l_6  , \, l_5  , \, l_4  , \, l_3  , \,  l_5-l_4   , \,  l_3-l_4   , \,  p_1-l_5  , \, 
    p_2-l_6   , \,  -l_4+l_5+l_6   , \,  -l_3+q   , \,  l_3-l_4+l_6-p_2  , \, \\  
    -l_3+l_4-l_5-l_6+q   , \,  l_3-l_5   , \,  l_3-l_6   , \,  l_5-l_6  , \, 
    l_4-p_1   , \,  l_4-p_2   , \,  l_5-p_2     \}  \,,
\end{multline}
\begin{multline}
\{   l_6  , \, l_5  , \, l_4  , \, l_3  , \,  l_5+l_6   , \,  p_2-l_6   , \,  p_1-l_4  , \, 
    l_3-l_4   , \,  -l_4-l_5+p_1   , \,  -l_3+q   , \,  -l_3+l_4+l_5+p_2  , \,
    -l_3+l_4+l_5+l_6   , \\    
    l_4-l_6   , \,  l_3-p_2   , \,  l_4-p_2  , \, 
    l_5-p_1   , \,  l_5-p_2   , \,  l_6-p_1     \}  \,,
\end{multline}
\begin{multline}
\{   l_6  , \, l_5  , \, l_4  , \, l_3  , \,  l_3-l_4   , \,  p_1-l_4   , \,  p_2-l_6  , \, 
    l_5-l_6   , \,  -l_3+q   , \,  l_3-l_4+l_5-p_2  , \,
    -l_3-l_5+q   , \,  -l_3+l_4-l_6+p_2   , \\    
    l_3-p_2   , \,  l_4-l_5  , \, 
    l_4-l_6   , \,  l_4-p_2   , \,  l_5-p_2   , \,  l_6-p_1     \}  \,,
\end{multline}
\begin{multline}
\{   l_6  , \, l_5  , \, l_4  , \, l_3  , \,  p_1-l_4   , \,  l_4-l_3   , \,  p_2-l_6  , \, 
    l_5+l_6   , \,  -l_4-l_5+p_1   , \,  -l_3+q   , \,  -l_3+l_4-l_6+p_2  , \, 
    -l_3-l_5-l_6+q   , \\    
    l_3-p_2   , \,  l_4-l_6   , \,  l_4-p_2  , \, 
    l_5-p_1   , \,  l_5-p_2   , \,  l_6-p_1     \}  \,,
\end{multline}
\begin{multline}
\{   l_6  , \, l_5  , \, l_4  , \, l_3  , \,  l_3-l_4   , \,  l_5+l_6   , \,  p_2-l_6  , \, 
    p_1-l_4   , \,  -l_3+q   , \,  -l_4-l_5+p_1   , \,  -l_3+l_4+l_5+p_2  , \,
    -l_3+l_4-l_6+p_2   , \\    
    l_3-p_2   , \,  l_4-l_6   , \,  l_4-p_2  , \, 
    l_5-p_1   , \,  l_5-p_2   , \,  l_6-p_1     \}  \,,
\end{multline}
\begin{multline}
\{   l_6  , \, l_5  , \, l_4  , \, l_3  , \,  l_3-p_1   , \,  p_2-l_6   , \,  l_5+l_6  , \, 
    l_4-l_5-p_2   , \,  -l_3-l_5+p_1   , \,  -l_3+q   , \,  l_4+l_6-p_2  , \,
    -l_3-l_4+q   , \\   
    l_3-l_6   , \,  l_4-p_1   , \,  l_4-p_2  , \, 
    l_5-p_1   , \,  l_5-p_2   , \,  l_6-p_1     \}  \,,
\end{multline}
\begin{multline}
\{   l_6  , \, l_5  , \, l_4  , \, l_3  , \,  l_3-p_1   , \,  p_2-l_6   , \,  l_5+l_6  , \, 
    -l_3-l_5+p_1   , \,  -l_4+l_5+p_2   , \,  -l_3+q   , \,  l_4-l_5-l_6  , \,
    -l_3-l_4+q   , \\    
    l_3-l_6   , \,  l_4-p_1   , \,  l_4-p_2  , \, 
    l_5-p_1   , \,  l_5-p_2   , \,  l_6-p_1     \}  \,,
\end{multline}
\begin{multline}
\{   l_6  , \, l_5  , \, l_4  , \, l_3  , \,  p_2-l_6   , \,  l_5+l_6   , \,  l_3-p_1  , \, 
    -l_3+q   , \,  l_4+l_6-p_2   , \,  -l_3-l_4+q  , \, 
    l_4+l_5+l_6-p_2   , \,  -l_3-l_4-l_5-l_6+q   , \\    
    l_3-l_6  , \, 
    l_4-p_1   , \,  l_4-p_2   , \,  l_5-p_1   , \,  l_5-p_2   , \,  l_6-p_1   \}  \,,
\end{multline}
\begin{multline}
\{   l_6  , \, l_5  , \, l_4  , \, l_3  , \,  p_2-l_6   , \,  l_3-p_1   , \,  l_5+l_6  , \, 
    l_4+l_6-p_2   , \,  -l_3-l_5+p_1   , \,  -l_3+q  , \, 
    -l_3-l_4+q   , \,  -l_3-l_4-l_5-l_6+q   , \\    
    l_3-l_6  , \,
    l_4-p_1   , \,  l_4-p_2   , \,  l_5-p_1   , \,  l_5-p_2   , \,  l_6-p_1    \}  \,.
\end{multline}
\abovedisplayskip=10pt
\belowdisplayskip=10pt

\section{List of ancillary files}

\begin{itemize}
\item List of master integrals appearing in the form factor obtained by {\tt Reduze}
\item List of master integrals as suggested by {\tt Mint}
\item IBP reduction of all integrals appearing in integral topology $25$ for the $\mathcal{N}=4$ SYM form factor
\end{itemize}

\bibliographystyle{elsarticle-num}

\bibliography{4LoopNPCuspNotes}

\begin{thebibliography}{10}
\expandafter\ifx\csname url\endcsname\relax
  \def\url#1{\texttt{#1}}\fi
\expandafter\ifx\csname urlprefix\endcsname\relax\def\urlprefix{URL }\fi
\expandafter\ifx\csname href\endcsname\relax
  \def\href#1#2{#2} \def\path#1{#1}\fi

\bibitem{Witten:2003nn}
E.~Witten, {Perturbative gauge theory as a string theory in twistor space},
  Commun. Math. Phys. 252 (2004) 189--258.
\newblock \href {http://arxiv.org/abs/hep-th/0312171}
  {\path{arXiv:hep-th/0312171}}, \href
  {http://dx.doi.org/10.1007/s00220-004-1187-3}
  {\path{doi:10.1007/s00220-004-1187-3}}.

\bibitem{vanNeerven:1985ja}
W.~van Neerven, {Infrared Behavior of On-shell Form-factors in a $N=4$
  Supersymmetric {Yang-Mills} Field Theory}, Z.Phys. C30 (1986) 595.
\newblock \href {http://dx.doi.org/10.1007/BF01571808}
  {\path{doi:10.1007/BF01571808}}.

\bibitem{Baikov:2009bg}
P.~Baikov, K.~Chetyrkin, A.~Smirnov, V.~Smirnov, M.~Steinhauser, {Quark and
  gluon form factors to three loops}, Phys.Rev.Lett. 102 (2009) 212002.
\newblock \href {http://arxiv.org/abs/0902.3519} {\path{arXiv:0902.3519}},
  \href {http://dx.doi.org/10.1103/PhysRevLett.102.212002}
  {\path{doi:10.1103/PhysRevLett.102.212002}}.

\bibitem{Gehrmann:2006wg}
T.~Gehrmann, G.~Heinrich, T.~Huber, C.~Studerus, {Master integrals for massless
  three-loop form-factors: One-loop and two-loop insertions}, Phys.Lett. B640
  (2006) 252--259.
\newblock \href {http://arxiv.org/abs/hep-ph/0607185}
  {\path{arXiv:hep-ph/0607185}}, \href
  {http://dx.doi.org/10.1016/j.physletb.2006.08.008}
  {\path{doi:10.1016/j.physletb.2006.08.008}}.

\bibitem{Gehrmann:2011xn}
T.~Gehrmann, J.~M. Henn, T.~Huber, {The three-loop form factor in N=4 super
  Yang-Mills}, JHEP 1203 (2012) 101.
\newblock \href {http://arxiv.org/abs/1112.4524} {\path{arXiv:1112.4524}},
  \href {http://dx.doi.org/10.1007/JHEP03(2012)101}
  {\path{doi:10.1007/JHEP03(2012)101}}.

\bibitem{Korchemsky:1985xj}
G.~P. Korchemsky, A.~V. Radyushkin, {Loop Space Formalism and Renormalization
  Group for the Infrared Asymptotics of {QCD}}, Phys. Lett. B171 (1986)
  459--467.
\newblock \href {http://dx.doi.org/10.1016/0370-2693(86)91439-5}
  {\path{doi:10.1016/0370-2693(86)91439-5}}.

\bibitem{Korchemsky:1987wg}
G.~P. Korchemsky, A.~V. Radyushkin, {Renormalization of the Wilson Loops Beyond
  the Leading Order}, Nucl. Phys. B283 (1987) 342--364.
\newblock \href {http://dx.doi.org/10.1016/0550-3213(87)90277-X}
  {\path{doi:10.1016/0550-3213(87)90277-X}}.

\bibitem{Beisert:2006ez}
N.~Beisert, B.~Eden, M.~Staudacher, {Transcendentality and Crossing},
  J.Stat.Mech. 0701 (2007) P01021.
\newblock \href {http://arxiv.org/abs/hep-th/0610251}
  {\path{arXiv:hep-th/0610251}}, \href
  {http://dx.doi.org/10.1088/1742-5468/2007/01/P01021}
  {\path{doi:10.1088/1742-5468/2007/01/P01021}}.

\bibitem{Maldacena:1997re}
J.~M. Maldacena, {The Large N limit of superconformal field theories and
  supergravity}, Int.J.Theor.Phys. 38 (1999) 1113--1133.
\newblock \href {http://arxiv.org/abs/hep-th/9711200}
  {\path{arXiv:hep-th/9711200}}, \href
  {http://dx.doi.org/10.1023/A:1026654312961}
  {\path{doi:10.1023/A:1026654312961}}.

\bibitem{Becher:2009qa}
T.~Becher, M.~Neubert, {On the Structure of Infrared Singularities of
  Gauge-Theory Amplitudes}, JHEP 0906 (2009) 081.
\newblock \href {http://arxiv.org/abs/0903.1126} {\path{arXiv:0903.1126}},
  \href {http://dx.doi.org/10.1088/1126-6708/2009/06/081,
  10.1007/JHEP11(2013)024} {\path{doi:10.1088/1126-6708/2009/06/081,
  10.1007/JHEP11(2013)024}}.

\bibitem{Gardi:2009qi}
E.~Gardi, L.~Magnea, {Factorization constraints for soft anomalous dimensions
  in QCD scattering amplitudes}, JHEP 03 (2009) 079.
\newblock \href {http://arxiv.org/abs/0901.1091} {\path{arXiv:0901.1091}},
  \href {http://dx.doi.org/10.1088/1126-6708/2009/03/079}
  {\path{doi:10.1088/1126-6708/2009/03/079}}.

\bibitem{Caron-Huot:2013fea}
S.~Caron-Huot, {When does the gluon reggeize?}, JHEP 1505 (2015) 093.
\newblock \href {http://arxiv.org/abs/1309.6521} {\path{arXiv:1309.6521}},
  \href {http://dx.doi.org/10.1007/JHEP05(2015)093}
  {\path{doi:10.1007/JHEP05(2015)093}}.

\bibitem{Boels:2012ew}
R.~H. Boels, B.~A. Kniehl, O.~V. Tarasov, G.~Yang, {Color-kinematic Duality for
  Form Factors}, JHEP 1302 (2013) 063.
\newblock \href {http://arxiv.org/abs/1211.7028} {\path{arXiv:1211.7028}},
  \href {http://dx.doi.org/10.1007/JHEP02(2013)063}
  {\path{doi:10.1007/JHEP02(2013)063}}.

\bibitem{Bern:2008qj}
Z.~Bern, J.~Carrasco, H.~Johansson, {New Relations for Gauge-Theory
  Amplitudes}, Phys.Rev. D78 (2008) 085011.
\newblock \href {http://arxiv.org/abs/0805.3993} {\path{arXiv:0805.3993}},
  \href {http://dx.doi.org/10.1103/PhysRevD.78.085011}
  {\path{doi:10.1103/PhysRevD.78.085011}}.

\bibitem{Bern:2010ue}
Z.~Bern, J.~J.~M. Carrasco, H.~Johansson, {Perturbative Quantum Gravity as a
  Double Copy of Gauge Theory}, Phys. Rev. Lett. 105 (2010) 061602.
\newblock \href {http://arxiv.org/abs/1004.0476} {\path{arXiv:1004.0476}},
  \href {http://dx.doi.org/10.1103/PhysRevLett.105.061602}
  {\path{doi:10.1103/PhysRevLett.105.061602}}.

\bibitem{Bern:2012uf}
Z.~Bern, J.~J.~M. Carrasco, L.~J. Dixon, H.~Johansson, R.~Roiban, {Simplifying
  Multiloop Integrands and Ultraviolet Divergences of Gauge Theory and Gravity
  Amplitudes}, Phys. Rev. D85 (2012) 105014.
\newblock \href {http://arxiv.org/abs/1201.5366} {\path{arXiv:1201.5366}},
  \href {http://dx.doi.org/10.1103/PhysRevD.85.105014}
  {\path{doi:10.1103/PhysRevD.85.105014}}.

\bibitem{Chetyrkin:1981qh}
K.~Chetyrkin, F.~Tkachov, {Integration by Parts: The Algorithm to Calculate
  beta Functions in 4 Loops}, Nucl.Phys. B192 (1981) 159--204.
\newblock \href {http://dx.doi.org/10.1016/0550-3213(81)90199-1}
  {\path{doi:10.1016/0550-3213(81)90199-1}}.

\bibitem{Tkachov:1981wb}
F.~Tkachov, {A Theorem on Analytical Calculability of Four Loop Renormalization
  Group Functions}, Phys.Lett. B100 (1981) 65--68.
\newblock \href {http://dx.doi.org/10.1016/0370-2693(81)90288-4}
  {\path{doi:10.1016/0370-2693(81)90288-4}}.

\bibitem{Smirnov:2012gma}
V.~A. Smirnov, {Analytic tools for Feynman integrals}, Springer Tracts
  Mod.Phys. 250 (2012) 1--296.
\newblock \href {http://dx.doi.org/10.1007/978-3-642-34886-0}
  {\path{doi:10.1007/978-3-642-34886-0}}.

\bibitem{vonManteuffel:2012np}
A.~von Manteuffel, C.~Studerus, {Reduze 2 - Distributed Feynman Integral
  Reduction}\href {http://arxiv.org/abs/1201.4330} {\path{arXiv:1201.4330}}.

\bibitem{Lee:2013hzt}
R.~N. Lee, A.~A. Pomeransky, {Critical points and number of master integrals},
  JHEP 1311 (2013) 165.
\newblock \href {http://arxiv.org/abs/1308.6676} {\path{arXiv:1308.6676}},
  \href {http://dx.doi.org/10.1007/JHEP11(2013)165}
  {\path{doi:10.1007/JHEP11(2013)165}}.

\bibitem{Brandhuber:2010ad}
A.~Brandhuber, B.~Spence, G.~Travaglini, G.~Yang, {Form Factors in N=4 Super
  Yang-Mills and Periodic Wilson Loops}, JHEP 01 (2011) 134.
\newblock \href {http://arxiv.org/abs/1011.1899} {\path{arXiv:1011.1899}},
  \href {http://dx.doi.org/10.1007/JHEP01(2011)134}
  {\path{doi:10.1007/JHEP01(2011)134}}.

\bibitem{Bork:2010wf}
L.~V. Bork, D.~I. Kazakov, G.~S. Vartanov, {On form factors in N=4 sym}, JHEP
  02 (2011) 063.
\newblock \href {http://arxiv.org/abs/1011.2440} {\path{arXiv:1011.2440}},
  \href {http://dx.doi.org/10.1007/JHEP02(2011)063}
  {\path{doi:10.1007/JHEP02(2011)063}}.

\bibitem{Brandhuber:2011tv}
A.~Brandhuber, O.~Gurdogan, R.~Mooney, G.~Travaglini, G.~Yang, {Harmony of
  Super Form Factors}, JHEP 10 (2011) 046.
\newblock \href {http://arxiv.org/abs/1107.5067} {\path{arXiv:1107.5067}},
  \href {http://dx.doi.org/10.1007/JHEP10(2011)046}
  {\path{doi:10.1007/JHEP10(2011)046}}.

\bibitem{Bork:2011cj}
L.~V. Bork, D.~I. Kazakov, G.~S. Vartanov, {On MHV Form Factors in Superspace
  for $\mathcal{n}=4$ SYM Theory}, JHEP 10 (2011) 133.
\newblock \href {http://arxiv.org/abs/1107.5551} {\path{arXiv:1107.5551}},
  \href {http://dx.doi.org/10.1007/JHEP10(2011)133}
  {\path{doi:10.1007/JHEP10(2011)133}}.

\bibitem{Henn:2011by}
J.~M. Henn, S.~Moch, S.~G. Naculich, {Form factors and scattering amplitudes in
  N=4 SYM in dimensional and massive regularizations}, JHEP 12 (2011) 024.
\newblock \href {http://arxiv.org/abs/1109.5057} {\path{arXiv:1109.5057}},
  \href {http://dx.doi.org/10.1007/JHEP12(2011)024}
  {\path{doi:10.1007/JHEP12(2011)024}}.

\bibitem{Brandhuber:2012vm}
A.~Brandhuber, G.~Travaglini, G.~Yang, {Analytic two-loop form factors in N=4
  SYM}, JHEP 05 (2012) 082.
\newblock \href {http://arxiv.org/abs/1201.4170} {\path{arXiv:1201.4170}},
  \href {http://dx.doi.org/10.1007/JHEP05(2012)082}
  {\path{doi:10.1007/JHEP05(2012)082}}.

\bibitem{Bork:2012tt}
L.~V. Bork, {On NMHV form factors in N=4 SYM theory from generalized
  unitarity}, JHEP 01 (2013) 049.
\newblock \href {http://arxiv.org/abs/1203.2596} {\path{arXiv:1203.2596}},
  \href {http://dx.doi.org/10.1007/JHEP01(2013)049}
  {\path{doi:10.1007/JHEP01(2013)049}}.

\bibitem{Engelund:2012re}
O.~T. Engelund, R.~Roiban, {Correlation functions of local composite operators
  from generalized unitarity}, JHEP 03 (2013) 172.
\newblock \href {http://arxiv.org/abs/1209.0227} {\path{arXiv:1209.0227}},
  \href {http://dx.doi.org/10.1007/JHEP03(2013)172}
  {\path{doi:10.1007/JHEP03(2013)172}}.

\bibitem{Penante:2014sza}
B.~Penante, B.~Spence, G.~Travaglini, C.~Wen, {On super form factors of
  half-BPS operators in N=4 super Yang-Mills}, JHEP 04 (2014) 083.
\newblock \href {http://arxiv.org/abs/1402.1300} {\path{arXiv:1402.1300}},
  \href {http://dx.doi.org/10.1007/JHEP04(2014)083}
  {\path{doi:10.1007/JHEP04(2014)083}}.

\bibitem{Brandhuber:2014ica}
A.~Brandhuber, B.~Penante, G.~Travaglini, C.~Wen, {The last of the simple
  remainders}, JHEP 08 (2014) 100.
\newblock \href {http://arxiv.org/abs/1406.1443} {\path{arXiv:1406.1443}},
  \href {http://dx.doi.org/10.1007/JHEP08(2014)100}
  {\path{doi:10.1007/JHEP08(2014)100}}.

\bibitem{Bork:2014eqa}
L.~V. Bork, {On form factors in $ \mathcal{N}=4 $ SYM theory and polytopes},
  JHEP 12 (2014) 111.
\newblock \href {http://arxiv.org/abs/1407.5568} {\path{arXiv:1407.5568}},
  \href {http://dx.doi.org/10.1007/JHEP12(2014)111}
  {\path{doi:10.1007/JHEP12(2014)111}}.

\bibitem{Alday:2007he}
L.~F. Alday, J.~Maldacena, {Comments on gluon scattering amplitudes via
  AdS/CFT}, JHEP 11 (2007) 068.
\newblock \href {http://arxiv.org/abs/0710.1060} {\path{arXiv:0710.1060}},
  \href {http://dx.doi.org/10.1088/1126-6708/2007/11/068}
  {\path{doi:10.1088/1126-6708/2007/11/068}}.

\bibitem{Maldacena:2010kp}
J.~Maldacena, A.~Zhiboedov, {Form factors at strong coupling via a Y-system},
  JHEP 11 (2010) 104.
\newblock \href {http://arxiv.org/abs/1009.1139} {\path{arXiv:1009.1139}},
  \href {http://dx.doi.org/10.1007/JHEP11(2010)104}
  {\path{doi:10.1007/JHEP11(2010)104}}.

\bibitem{Gao:2013dza}
Z.~Gao, G.~Yang, {Y-system for form factors at strong coupling in $AdS_5$ and
  with multi-operator insertions in $AdS_3$}, JHEP 1306 (2013) 105.
\newblock \href {http://arxiv.org/abs/1303.2668} {\path{arXiv:1303.2668}},
  \href {http://dx.doi.org/10.1007/JHEP06(2013)105}
  {\path{doi:10.1007/JHEP06(2013)105}}.

\bibitem{Wilhelm:2014qua}
M.~Wilhelm, {Amplitudes, Form Factors and the Dilatation Operator in
  $\mathcal{N}=4$ SYM Theory}, JHEP 02 (2015) 149.
\newblock \href {http://arxiv.org/abs/1410.6309} {\path{arXiv:1410.6309}},
  \href {http://dx.doi.org/10.1007/JHEP02(2015)149}
  {\path{doi:10.1007/JHEP02(2015)149}}.

\bibitem{Nandan:2014oga}
D.~Nandan, C.~Sieg, M.~Wilhelm, G.~Yang, {Cutting through form factors and
  cross sections of non-protected operators in $ \mathcal{N}=4 $ SYM}, JHEP 06
  (2015) 156.
\newblock \href {http://arxiv.org/abs/1410.8485} {\path{arXiv:1410.8485}},
  \href {http://dx.doi.org/10.1007/JHEP06(2015)156}
  {\path{doi:10.1007/JHEP06(2015)156}}.

\bibitem{Loebbert:2015ova}
F.~Loebbert, D.~Nandan, C.~Sieg, M.~Wilhelm, G.~Yang, {On-Shell Methods for the
  Two-Loop Dilatation Operator and Finite Remainders}, JHEP 10 (2015) 012.
\newblock \href {http://arxiv.org/abs/1504.06323} {\path{arXiv:1504.06323}},
  \href {http://dx.doi.org/10.1007/JHEP10(2015)012}
  {\path{doi:10.1007/JHEP10(2015)012}}.

\bibitem{Frassek:2015rka}
R.~Frassek, D.~Meidinger, D.~Nandan, M.~Wilhelm, {On-shell Diagrams,
  Gra{\ss}mannians and Integrability for Form Factors}\href
  {http://arxiv.org/abs/1506.08192} {\path{arXiv:1506.08192}}.

\bibitem{Mueller:1979ih}
A.~H. Mueller, {On the Asymptotic Behavior of the Sudakov Form-factor},
  Phys.Rev. D20 (1979) 2037.
\newblock \href {http://dx.doi.org/10.1103/PhysRevD.20.2037}
  {\path{doi:10.1103/PhysRevD.20.2037}}.

\bibitem{Collins:1980ih}
J.~C. Collins, {Algorithm to Compute Corrections to the Sudakov Form-factor},
  Phys.Rev. D22 (1980) 1478.
\newblock \href {http://dx.doi.org/10.1103/PhysRevD.22.1478}
  {\path{doi:10.1103/PhysRevD.22.1478}}.

\bibitem{Sen:1981sd}
A.~Sen, {Asymptotic Behavior of the Sudakov Form-Factor in QCD}, Phys.Rev. D24
  (1981) 3281.
\newblock \href {http://dx.doi.org/10.1103/PhysRevD.24.3281}
  {\path{doi:10.1103/PhysRevD.24.3281}}.

\bibitem{Magnea:1990zb}
L.~Magnea, G.~F. Sterman, {Analytic continuation of the Sudakov form-factor in
  QCD}, Phys.Rev. D42 (1990) 4222--4227.
\newblock \href {http://dx.doi.org/10.1103/PhysRevD.42.4222}
  {\path{doi:10.1103/PhysRevD.42.4222}}.

\bibitem{Bern:2005iz}
Z.~Bern, L.~J. Dixon, V.~A. Smirnov, {Iteration of planar amplitudes in
  maximally supersymmetric Yang-Mills theory at three loops and beyond}, Phys.
  Rev. D72 (2005) 085001.
\newblock \href {http://arxiv.org/abs/hep-th/0505205}
  {\path{arXiv:hep-th/0505205}}, \href
  {http://dx.doi.org/10.1103/PhysRevD.72.085001}
  {\path{doi:10.1103/PhysRevD.72.085001}}.

\bibitem{Bern:2006ew}
Z.~Bern, M.~Czakon, L.~J. Dixon, D.~A. Kosower, V.~A. Smirnov, {The Four-Loop
  Planar Amplitude and Cusp Anomalous Dimension in Maximally Supersymmetric
  Yang-Mills Theory}, Phys. Rev. D75 (2007) 085010.
\newblock \href {http://arxiv.org/abs/hep-th/0610248}
  {\path{arXiv:hep-th/0610248}}, \href
  {http://dx.doi.org/10.1103/PhysRevD.75.085010}
  {\path{doi:10.1103/PhysRevD.75.085010}}.

\bibitem{Cachazo:2006az}
F.~Cachazo, M.~Spradlin, A.~Volovich, {Four-loop cusp anomalous dimension from
  obstructions}, Phys. Rev. D75 (2007) 105011.
\newblock \href {http://arxiv.org/abs/hep-th/0612309}
  {\path{arXiv:hep-th/0612309}}, \href
  {http://dx.doi.org/10.1103/PhysRevD.75.105011}
  {\path{doi:10.1103/PhysRevD.75.105011}}.

\bibitem{Henn:2013wfa}
J.~M. Henn, T.~Huber, {The four-loop cusp anomalous dimension in $\mathcal{N}
  =$ 4 super Yang-Mills and analytic integration techniques for Wilson line
  integrals}, JHEP 09 (2013) 147.
\newblock \href {http://arxiv.org/abs/1304.6418} {\path{arXiv:1304.6418}},
  \href {http://dx.doi.org/10.1007/JHEP09(2013)147}
  {\path{doi:10.1007/JHEP09(2013)147}}.

\bibitem{Kotikov:2004er}
A.~V. Kotikov, L.~N. Lipatov, A.~I. Onishchenko, V.~N. Velizhanin, {Three loop
  universal anomalous dimension of the Wilson operators in N=4 SUSY Yang-Mills
  model}, Phys. Lett. B595 (2004) 521--529, [Erratum: Phys.
  Lett.B632,754(2006)].
\newblock \href {http://arxiv.org/abs/hep-th/0404092}
  {\path{arXiv:hep-th/0404092}}, \href
  {http://dx.doi.org/10.1016/j.physletb.2004.05.078}
  {\path{doi:10.1016/j.physletb.2004.05.078}}.

\bibitem{Moch:2004pa}
S.~Moch, J.~A.~M. Vermaseren, A.~Vogt, {The Three loop splitting functions in
  QCD: The Nonsinglet case}, Nucl. Phys. B688 (2004) 101--134.
\newblock \href {http://arxiv.org/abs/hep-ph/0403192}
  {\path{arXiv:hep-ph/0403192}}, \href
  {http://dx.doi.org/10.1016/j.nuclphysb.2004.03.030}
  {\path{doi:10.1016/j.nuclphysb.2004.03.030}}.

\bibitem{Vogt:2004mw}
A.~Vogt, S.~Moch, J.~A.~M. Vermaseren, {The Three-loop splitting functions in
  QCD: The Singlet case}, Nucl. Phys. B691 (2004) 129--181.
\newblock \href {http://arxiv.org/abs/hep-ph/0404111}
  {\path{arXiv:hep-ph/0404111}}, \href
  {http://dx.doi.org/10.1016/j.nuclphysb.2004.04.024}
  {\path{doi:10.1016/j.nuclphysb.2004.04.024}}.

\bibitem{'tHooft:1973jz}
G.~'t~Hooft, {A Planar Diagram Theory for Strong Interactions}, Nucl.Phys. B72
  (1974) 461.
\newblock \href {http://dx.doi.org/10.1016/0550-3213(74)90154-0}
  {\path{doi:10.1016/0550-3213(74)90154-0}}.

\bibitem{asmirnovwebsite}
A.~V. Smirnov, \url{https://science.sander.su/Tools.htm}.

\bibitem{Pak:2011xt}
A.~Pak, {The Toolbox of modern multi-loop calculations: novel analytic and
  semi-analytic techniques}, J.Phys.Conf.Ser. 368 (2012) 012049.
\newblock \href {http://arxiv.org/abs/1111.0868} {\path{arXiv:1111.0868}},
  \href {http://dx.doi.org/10.1088/1742-6596/368/1/012049}
  {\path{doi:10.1088/1742-6596/368/1/012049}}.

\bibitem{Laporta:2001dd}
S.~Laporta, {High precision calculation of multiloop Feynman integrals by
  difference equations}, Int.J.Mod.Phys. A15 (2000) 5087--5159.
\newblock \href {http://arxiv.org/abs/hep-ph/0102033}
  {\path{arXiv:hep-ph/0102033}}, \href
  {http://dx.doi.org/10.1016/S0217-751X(00)00215-7}
  {\path{doi:10.1016/S0217-751X(00)00215-7}}.

\bibitem{Anastasiou:2004vj}
C.~Anastasiou, A.~Lazopoulos, {Automatic integral reduction for higher order
  perturbative calculations}, JHEP 0407 (2004) 046.
\newblock \href {http://arxiv.org/abs/hep-ph/0404258}
  {\path{arXiv:hep-ph/0404258}}, \href
  {http://dx.doi.org/10.1088/1126-6708/2004/07/046}
  {\path{doi:10.1088/1126-6708/2004/07/046}}.

\bibitem{Smirnov:2008iw}
A.~Smirnov, {Algorithm FIRE -- Feynman Integral REduction}, JHEP 0810 (2008)
  107.
\newblock \href {http://arxiv.org/abs/0807.3243} {\path{arXiv:0807.3243}},
  \href {http://dx.doi.org/10.1088/1126-6708/2008/10/107}
  {\path{doi:10.1088/1126-6708/2008/10/107}}.

\bibitem{Smirnov:2013dia}
A.~Smirnov, V.~Smirnov, {FIRE4, LiteRed and accompanying tools to solve
  integration by parts relations}, Comput.Phys.Commun. 184 (2013) 2820--2827.
\newblock \href {http://arxiv.org/abs/1302.5885} {\path{arXiv:1302.5885}},
  \href {http://dx.doi.org/10.1016/j.cpc.2013.06.016}
  {\path{doi:10.1016/j.cpc.2013.06.016}}.

\bibitem{Smirnov:2014hma}
A.~V. Smirnov, {FIRE5: a C++ implementation of Feynman Integral REduction},
  Comput.Phys.Commun. 189 (2014) 182--191.
\newblock \href {http://arxiv.org/abs/1408.2372} {\path{arXiv:1408.2372}},
  \href {http://dx.doi.org/10.1016/j.cpc.2014.11.024}
  {\path{doi:10.1016/j.cpc.2014.11.024}}.

\bibitem{2010CoPhC.181.1293S}
C.~{Studerus}, {Reduze - Feynman integral reduction in C++},
  Comput.Phys.Commun. 181 (2010) 1293--1300.
\newblock \href {http://arxiv.org/abs/0912.2546} {\path{arXiv:0912.2546}},
  \href {http://dx.doi.org/10.1016/j.cpc.2010.03.012}
  {\path{doi:10.1016/j.cpc.2010.03.012}}.

\bibitem{Lee:2012cn}
R.~Lee, {Presenting LiteRed: a tool for the Loop InTEgrals REDuction}\href
  {http://arxiv.org/abs/1212.2685} {\path{arXiv:1212.2685}}.

\bibitem{Lee:2013mka}
R.~N. Lee, {LiteRed 1.4: a powerful tool for reduction of multiloop integrals},
  J.Phys.Conf.Ser. 523 (2014) 012059.
\newblock \href {http://arxiv.org/abs/1310.1145} {\path{arXiv:1310.1145}},
  \href {http://dx.doi.org/10.1088/1742-6596/523/1/012059}
  {\path{doi:10.1088/1742-6596/523/1/012059}}.

\bibitem{Hammond:2014:IBE:2690883.2690884}
J.~R. Hammond, A.~Sch\"{a}fer, R.~Latham,
  \href{http://dx.doi.org/10.1109/ExaMPI.2014.5}{To int\_max... and beyond!:
  Exploring large-count support in mpi}, in: Proceedings of the 2014 Workshop
  on Exascale MPI, ExaMPI '14, IEEE Press, Piscataway, NJ, USA, 2014, pp. 1--8.
\newblock \href {http://dx.doi.org/10.1109/ExaMPI.2014.5}
  {\path{doi:10.1109/ExaMPI.2014.5}}.
\newline\urlprefix\url{http://dx.doi.org/10.1109/ExaMPI.2014.5}

\bibitem{Baikov:2005nv}
P.~A. Baikov, {A Practical criterion of irreducibility of multi-loop Feynman
  integrals}, Phys. Lett. B634 (2006) 325--329.
\newblock \href {http://arxiv.org/abs/hep-ph/0507053}
  {\path{arXiv:hep-ph/0507053}}, \href
  {http://dx.doi.org/10.1016/j.physletb.2006.01.052}
  {\path{doi:10.1016/j.physletb.2006.01.052}}.

\bibitem{cox2007ideals}
D.~A. Cox, J.~Little, D.~OSHEA, Ideals, varieties, and algorithms: an
  introduction to computational algebraic geometry and commutative algebra,
  Springer Science \& Business Media, 2007.

\bibitem{M2}
D.~R. Grayson, M.~E. Stillman, Macaulay2, a software system for research in
  algebraic geometry, Available at \href{http://www.math.uiuc.edu/Macaulay2/}%
  {http://www.math.uiuc.edu/Macaulay2/}.

\bibitem{mathematicaM2}
Y.~Zhang, Interface to macaulay 2 from mathemtica,
  \url{https://bitbucket.org/yzhphy/mathematicam2} (2015).

\bibitem{DGPS}
W.~Decker, G.-M. Greuel, G.~Pfister, H.~Sch\"onemann, {\sc Singular} {4-0-2}
  --- {A} computer algebra system for polynomial computations,
  \url{http://www.singular.uni-kl.de} (2015).

\bibitem{smirnov-private}
V.~A. Smirnov, {Private communication}.

\bibitem{Czakon:2005rk}
M.~Czakon, {Automatized analytic continuation of Mellin-Barnes integrals},
  Comput.Phys.Commun. 175 (2006) 559--571.
\newblock \href {http://arxiv.org/abs/hep-ph/0511200}
  {\path{arXiv:hep-ph/0511200}}, \href
  {http://dx.doi.org/10.1016/j.cpc.2006.07.002}
  {\path{doi:10.1016/j.cpc.2006.07.002}}.

\bibitem{Smirnov:2009up}
A.~V. Smirnov, V.~A. Smirnov, {On the Resolution of Singularities of Multiple
  Mellin-Barnes Integrals}, Eur. Phys. J. C62 (2009) 445--449.
\newblock \href {http://arxiv.org/abs/0901.0386} {\path{arXiv:0901.0386}},
  \href {http://dx.doi.org/10.1140/epjc/s10052-009-1039-6}
  {\path{doi:10.1140/epjc/s10052-009-1039-6}}.

\bibitem{Gluza:2007rt}
J.~Gluza, K.~Kajda, T.~Riemann, {AMBRE: A Mathematica package for the
  construction of Mellin-Barnes representations for Feynman integrals},
  Comput.Phys.Commun. 177 (2007) 879--893.
\newblock \href {http://arxiv.org/abs/0704.2423} {\path{arXiv:0704.2423}},
  \href {http://dx.doi.org/10.1016/j.cpc.2007.07.001}
  {\path{doi:10.1016/j.cpc.2007.07.001}}.

\bibitem{Gluza:2010rn}
J.~Gluza, K.~Kajda, T.~Riemann, V.~Yundin, {Numerical Evaluation of Tensor
  Feynman Integrals in Euclidean Kinematics}, Eur.Phys.J. C71 (2011) 1516.
\newblock \href {http://arxiv.org/abs/1010.1667} {\path{arXiv:1010.1667}},
  \href {http://dx.doi.org/10.1140/epjc/s10052-010-1516-y}
  {\path{doi:10.1140/epjc/s10052-010-1516-y}}.

\bibitem{Blumlein:2014maa}
J.~Blümlein, I.~Dubovyk, J.~Gluza, M.~Ochman, C.~G. Raab, et~al., {Non-planar
  Feynman integrals, Mellin-Barnes representations, multiple sums}, PoS LL2014
  (2014) 052.
\newblock \href {http://arxiv.org/abs/1407.7832} {\path{arXiv:1407.7832}}.

\bibitem{Binoth:2000ps}
T.~Binoth, G.~Heinrich, {An automatized algorithm to compute infrared divergent
  multiloop integrals}, Nucl.Phys. B585 (2000) 741--759.
\newblock \href {http://arxiv.org/abs/hep-ph/0004013}
  {\path{arXiv:hep-ph/0004013}}, \href
  {http://dx.doi.org/10.1016/S0550-3213(00)00429-6}
  {\path{doi:10.1016/S0550-3213(00)00429-6}}.

\bibitem{Hahn:2004fe}
T.~Hahn, {CUBA: A Library for multidimensional numerical integration},
  Comput.Phys.Commun. 168 (2005) 78--95.
\newblock \href {http://arxiv.org/abs/hep-ph/0404043}
  {\path{arXiv:hep-ph/0404043}}, \href
  {http://dx.doi.org/10.1016/j.cpc.2005.01.010}
  {\path{doi:10.1016/j.cpc.2005.01.010}}.

\bibitem{Bogner:2007cr}
C.~Bogner, S.~Weinzierl, {Resolution of singularities for multi-loop
  integrals}, Comput.Phys.Commun. 178 (2008) 596--610.
\newblock \href {http://arxiv.org/abs/0709.4092} {\path{arXiv:0709.4092}},
  \href {http://dx.doi.org/10.1016/j.cpc.2007.11.012}
  {\path{doi:10.1016/j.cpc.2007.11.012}}.

\bibitem{Carter:2010hi}
J.~Carter, G.~Heinrich, {SecDec: A general program for sector decomposition},
  Comput.Phys.Commun. 182 (2011) 1566--1581.
\newblock \href {http://arxiv.org/abs/1011.5493} {\path{arXiv:1011.5493}},
  \href {http://dx.doi.org/10.1016/j.cpc.2011.03.026}
  {\path{doi:10.1016/j.cpc.2011.03.026}}.

\bibitem{Borowka:2012yc}
S.~Borowka, J.~Carter, G.~Heinrich, {Numerical Evaluation of Multi-Loop
  Integrals for Arbitrary Kinematics with SecDec 2.0}, Comput.Phys.Commun. 184
  (2013) 396--408.
\newblock \href {http://arxiv.org/abs/1204.4152} {\path{arXiv:1204.4152}},
  \href {http://dx.doi.org/10.1016/j.cpc.2012.09.020}
  {\path{doi:10.1016/j.cpc.2012.09.020}}.

\bibitem{Borowka:2015mxa}
S.~Borowka, G.~Heinrich, S.~P. Jones, M.~Kerner, J.~Schlenk, T.~Zirke,
  {SecDec-3.0: numerical evaluation of multi-scale integrals beyond one loop},
  Comput. Phys. Commun. 196 (2015) 470--491.
\newblock \href {http://arxiv.org/abs/1502.06595} {\path{arXiv:1502.06595}},
  \href {http://dx.doi.org/10.1016/j.cpc.2015.05.022}
  {\path{doi:10.1016/j.cpc.2015.05.022}}.

\bibitem{Smirnov:2008py}
A.~Smirnov, M.~Tentyukov, {Feynman Integral Evaluation by a Sector
  decomposiTion Approach (FIESTA)}, Comput.Phys.Commun. 180 (2009) 735--746.
\newblock \href {http://arxiv.org/abs/0807.4129} {\path{arXiv:0807.4129}},
  \href {http://dx.doi.org/10.1016/j.cpc.2008.11.006}
  {\path{doi:10.1016/j.cpc.2008.11.006}}.

\bibitem{Smirnov:2009pb}
A.~Smirnov, V.~Smirnov, M.~Tentyukov, {FIESTA 2: Parallelizeable multiloop
  numerical calculations}, Comput.Phys.Commun. 182 (2011) 790--803.
\newblock \href {http://arxiv.org/abs/0912.0158} {\path{arXiv:0912.0158}},
  \href {http://dx.doi.org/10.1016/j.cpc.2010.11.025}
  {\path{doi:10.1016/j.cpc.2010.11.025}}.

\bibitem{Smirnov:2013eza}
A.~V. Smirnov, {FIESTA 3: cluster-parallelizable multiloop numerical
  calculations in physical regions}, Comput.Phys.Commun. 185 (2014) 2090--2100.
\newblock \href {http://arxiv.org/abs/1312.3186} {\path{arXiv:1312.3186}},
  \href {http://dx.doi.org/10.1016/j.cpc.2014.03.015}
  {\path{doi:10.1016/j.cpc.2014.03.015}}.

\bibitem{Kaneko:2009qx}
T.~Kaneko, T.~Ueda, {A Geometric method of sector decomposition},
  Comput.Phys.Commun. 181 (2010) 1352--1361.
\newblock \href {http://arxiv.org/abs/0908.2897} {\path{arXiv:0908.2897}},
  \href {http://dx.doi.org/10.1016/j.cpc.2010.04.001}
  {\path{doi:10.1016/j.cpc.2010.04.001}}.

\bibitem{Lee:2010cga}
R.~Lee, A.~Smirnov, V.~Smirnov, {Analytic Results for Massless Three-Loop Form
  Factors}, JHEP 1004 (2010) 020.
\newblock \href {http://arxiv.org/abs/1001.2887} {\path{arXiv:1001.2887}},
  \href {http://dx.doi.org/10.1007/JHEP04(2010)020}
  {\path{doi:10.1007/JHEP04(2010)020}}.

\bibitem{Gehrmann:2010ue}
T.~Gehrmann, E.~Glover, T.~Huber, N.~Ikizlerli, C.~Studerus, {Calculation of
  the quark and gluon form factors to three loops in QCD}, JHEP 1006 (2010)
  094.
\newblock \href {http://arxiv.org/abs/1004.3653} {\path{arXiv:1004.3653}},
  \href {http://dx.doi.org/10.1007/JHEP06(2010)094}
  {\path{doi:10.1007/JHEP06(2010)094}}.

\bibitem{Henn:2013nsa}
J.~M. Henn, A.~V. Smirnov, V.~A. Smirnov, {Evaluating single-scale and/or
  non-planar diagrams by differential equations}, JHEP 1403 (2014) 088.
\newblock \href {http://arxiv.org/abs/1312.2588} {\path{arXiv:1312.2588}},
  \href {http://dx.doi.org/10.1007/JHEP03(2014)088}
  {\path{doi:10.1007/JHEP03(2014)088}}.

\bibitem{vonManteuffel:2014qoa}
A.~von Manteuffel, E.~Panzer, R.~M. Schabinger, {A quasi-finite basis for
  multi-loop Feynman integrals}, JHEP 1502 (2015) 120.
\newblock \href {http://arxiv.org/abs/1411.7392} {\path{arXiv:1411.7392}},
  \href {http://dx.doi.org/10.1007/JHEP02(2015)120}
  {\path{doi:10.1007/JHEP02(2015)120}}.

\bibitem{Tarasov:1996br}
O.~Tarasov, {Connection between Feynman integrals having different values of
  the space-time dimension}, Phys.Rev. D54 (1996) 6479--6490.
\newblock \href {http://arxiv.org/abs/hep-th/9606018}
  {\path{arXiv:hep-th/9606018}}, \href
  {http://dx.doi.org/10.1103/PhysRevD.54.6479}
  {\path{doi:10.1103/PhysRevD.54.6479}}.

\bibitem{Lee:2009dh}
R.~Lee, {Space-time dimensionality D as complex variable: Calculating loop
  integrals using dimensional recurrence relation and analytical properties
  with respect to D}, Nucl.Phys. B830 (2010) 474--492.
\newblock \href {http://arxiv.org/abs/0911.0252} {\path{arXiv:0911.0252}},
  \href {http://dx.doi.org/10.1016/j.nuclphysb.2009.12.025}
  {\path{doi:10.1016/j.nuclphysb.2009.12.025}}.

\bibitem{vonmanteuffelradcor}
A.~von Manteuffel, Talk at radcor/loopfest 2015, 2015.

\bibitem{vonManteuffel:2014ixa}
A.~von Manteuffel, R.~M. Schabinger, {A novel approach to integration by parts
  reduction}, Phys. Lett. B744 (2015) 101--104.
\newblock \href {http://arxiv.org/abs/1406.4513} {\path{arXiv:1406.4513}},
  \href {http://dx.doi.org/10.1016/j.physletb.2015.03.029}
  {\path{doi:10.1016/j.physletb.2015.03.029}}.

\bibitem{Smirnov:1999wz}
V.~A. Smirnov, O.~Veretin, {Analytical results for dimensionally regularized
  massless on-shell double boxes with arbitrary indices and numerators},
  Nucl.Phys. B566 (2000) 469--485.
\newblock \href {http://arxiv.org/abs/hep-ph/9907385}
  {\path{arXiv:hep-ph/9907385}}, \href
  {http://dx.doi.org/10.1016/S0550-3213(99)00686-0}
  {\path{doi:10.1016/S0550-3213(99)00686-0}}.

\bibitem{Anastasiou:2000kp}
C.~Anastasiou, J.~Tausk, M.~Tejeda-Yeomans, {The On-shell massless planar
  double box diagram with an irreducible numerator}, Nucl.Phys.Proc.Suppl. 89
  (2000) 262--267.
\newblock \href {http://arxiv.org/abs/hep-ph/0005328}
  {\path{arXiv:hep-ph/0005328}}, \href
  {http://dx.doi.org/10.1016/S0920-5632(00)00853-7}
  {\path{doi:10.1016/S0920-5632(00)00853-7}}.

\bibitem{Panzer:2014caa}
E.~Panzer, {Algorithms for the symbolic integration of hyperlogarithms with
  applications to Feynman integrals}, Comput. Phys. Commun. 188 (2014)
  148--166.
\newblock \href {http://arxiv.org/abs/1403.3385} {\path{arXiv:1403.3385}},
  \href {http://dx.doi.org/10.1016/j.cpc.2014.10.019}
  {\path{doi:10.1016/j.cpc.2014.10.019}}.

\bibitem{Henn:2013pwa}
J.~M. Henn, {Multiloop integrals in dimensional regularization made simple},
  Phys.Rev.Lett. 110~(25) (2013) 251601.
\newblock \href {http://arxiv.org/abs/1304.1806} {\path{arXiv:1304.1806}},
  \href {http://dx.doi.org/10.1103/PhysRevLett.110.251601}
  {\path{doi:10.1103/PhysRevLett.110.251601}}.

\end{thebibliography}

\end{document}